\documentclass{amsart}
\usepackage{amssymb}
\usepackage{bbm}
\usepackage{graphicx} 		% For postscript
\usepackage{epic,eepic}       % For epic and eepic output from xfig
\usepackage[foot]{amsaddr}
\usepackage[a4paper, top=3cm, left=4cm, right=4cm, bottom=3cm]{geometry}

\theoremstyle{definition}

\theoremstyle{remark}

\numberwithin{equation}{section}
\usepackage{hyperref}  
\hypersetup{
linktocpage,
colorlinks=true
,urlcolor=blue
,anchorcolor=blue
,citecolor=blue
,filecolor=blue
,linkcolor=blue
,menucolor=blue
}% link references
\usepackage[style=numeric,sorting=none]{biblatex} % bibliography
\usepackage{caption}
\usepackage{subcaption}                 % for subfigures in figures
\usepackage{cleveref}

\begin{document}

%% The title of the paper goes here.  Edit to your title.
%%

\title{Algebraic Structure of Dirac Hamiltonians in Non-Commutative Phase Space}

%%
%% Now edit the following to give your name and address:
%% 

\author{Horacio Falomir$^1$}
\address{$^1$Instituto de Física La Plata
(CONICET and Universidad Nacional de La Plata)
CC 67 (1900) La Plata, Argentina}
\email{falomir@fisica.unlp.edu.ar}
% Delete if not wanted.

\author{Joaquin Liniado$^1$}
\email{jliniado@iflp.unlp.edu.ar}
 % Delete if not wanted.

\author{Pablo Pisani$^1$}
\email{pisani@fisica.unlp.edu.ar}
 % Delete if not wanted.

%%
%% If there is another author uncomment and edit the following.
%%

%\author{Second Author}
%\address{Department of Mathematics, University of South Carolina,
%Columbia, SC 29208}
%\email{second@math.sc.edu}
%\urladdr{www.math.sc.edu/$\sim$second}

%%
%% If there are three of more authors they are added in the obvious
%% way. 
%%

%%%
%%% The following is for the abstract.  The abstract is optional and
%%% if not used just delete, or comment out, the following.
%%%

%%
%%  LaTeX will not make the title for the paper unless told to do so.
%%  This is done by uncommenting the following.
%%

% \maketitle

%%
%% LaTeX can automatically make a table of contents.  This is done by
%% uncommenting the following:
%%

%\tableofcontents

\begin{abstract}
In this article we study two-dimensional Dirac Hamiltonians with non-commutativity both in coordinates and momenta from an algebraic perspective. In order to do so, we consider the graded Lie algebra $\mathfrak{sl}(2|1)$ generated by Hermitian bilinear forms in the non-commutative dynamical variables and the Dirac matrices in $2+1$ dimensions. By further defining a total angular momentum operator, we are able to express a class of Dirac Hamiltonians completely in terms of these operators. In this way, we analyze the energy spectrum of some simple models by constructing and studying the representation spaces of the unitary irreducible representations of the graded Lie algebra $\mathfrak{sl}(2|1)\oplus \mathfrak{so}(2)$. As application of our results, we consider the Landau model and a fermion in a finite cylindrical well.
\end{abstract}

\maketitle

\newpage

\tableofcontents
% ----------------------------------------------------------------

\pagestyle{plain}

\section{Introduction}

Non-commutativity (NC) in physics is an area of research that has provided different models with peculiar properties, ever since its appearance in the 1930s' in the seminal paper by Peierls \cite{Peierls} (see also \cite{Heisenberg1,Heisenberg}), describing non-relativistic electrons subject to a strong magnetic field. Non-commutative geometries \cite{Connes} have been considered in a wide range of settings, originally in the context of regularization schemes in quantum field theory \cite{Snyder1,Snyder2,Yang}, as well as in string theory, in the description of low energy effective theories of $D-$branes in background magnetic fields \cite{Seiberg}. From a slightly different perspective, non-commutativity has also stimulated interest in the study of quantum mechanical systems with deformed algebra of commutators \cite{Dunne, Duval:2000xr, Mezincescu, Nair, Gamboa, Duval:2001hu,Douglas,Bellucci, Szabo:2001kg, Horvathy, Alvarez, Plyushchay, Panella:2014hga}, since these present notable properties either because of their phenomenological consequences or as an arena for the use of alternative techniques (\emph{cf.} \cite{Falomir1} for a detailed historical review and a list of references).

Non-commutativity has been implemented through the Moyal product of ordinary functions \cite{Mezincescu, Douglas, Szabo:2001kg} or, alternatively, by explicit realizations of dynamical variables as operators on the Hilbert space satisfying a deformed Heisenberg algebra of
the form \cite{Duval:2000xr, Mezincescu, Nair}

\begin{equation}
\label{ec:heisenbergedeform}
    \left[\hat{X}_i,\hat{X}_j\right]=i \theta \epsilon_{ij}\,, \quad   \left[\hat{X}_i,\hat{P}_j\right]=i \delta_{ij}\,,\quad   \left[\hat{P}_i,\hat{P}_j\right]=i \kappa \epsilon_{ij} \,,
\end{equation}
where $\kappa,\theta$ are real parameters describing phase-space NC, and we take $\hbar=1$. Several authors have pointed out that non-relativistic quantum mechanical models on the non-commutative phase space present two quantum phases \cite{Duval:2000xr, Nair, Bellucci} --according to the value of the dimensionless parameter $\kappa\theta$-- separated by a critical point where the effective dimension of the system is reduced \cite{Duval:2000xr, Nair, Bellucci, Duval:2001hu,  Horvathy,Alvarez}. In particular, for the case of Dirac Hamiltonians see \cite{Plyushchay, Panella:2014hga}.

In this context a class of non-relativistic two-dimensional rotationally invariant Hamiltonians have been studied from an algebraic point of view \cite{Falomir1}. Making use of the commutation relations \eqref{ec:heisenbergedeform}, it has been shown that the rotationally invariant \emph{Hermitian quadratic forms} in $\hat{X}_i,\hat{P}_i$, $i = 1, 2$ generate, in particular, a non-abelian three-dimensional Lie algebra corresponding to $\mathfrak{sl}(2,\mathbb{R})$ or $\mathfrak{su}(2)$ according to $\kappa \theta <1$ or $\kappa \theta >1$. By further constructing the generator of rotations $\hat{L}$ in the non-commutative phase space, it was established that the most general rotationally invariant non-relativistic Hamiltonian can be expressed as a function of the generators of that Lie algebra and $\hat{L}$ itself, so that the characteristic subspaces of these Hamiltonians are representation spaces of the unitary irreducible representations (irreps) of the Lie algebras $\mathfrak{sl}(2,\mathbb{R}) \oplus \mathfrak{so}(2)$ or $\mathfrak{su}(2) \oplus \mathfrak{so}(2)$. 

The aim of this article is to extend this algebraic study to two-dimensional \emph{fermionic} systems: we study Dirac Hamiltonians in the non-commutative phase space. We show that due to the spinorial structure of Dirac equation, the natural generalization is to consider a graded extension of $\mathfrak{sl}(2,\mathbb{R})$, the Lie algebra appearing\footnote{For simplicity, we restrict to the parameter region $\kappa\theta < 1$ for it is the one which can be connected with the ordinary (commutative) case. As mentioned above, the Lie algebra corresponding to this region is precisely $\mathfrak{sl}(2,\mathbb{R})$.} in the Schr\"odinger case \cite{Falomir1}. In fact, by constructing a \textit{total} angular momentum operator $\mathcal{L} $, we manage to express some Dirac-type Hamiltonians in terms of $\mathcal{L}$ and the generators of the graded Lie algebra $\mathfrak{sl}(2|1)$, which we then employ to compute, in two examples, the energy spectrum by studying, as in the previous case, the representation spaces of the irreducible representations of the graded Lie algebra. Specifically, we discuss the non-commutative version of an electron subject to a uniform magnetic field (the NC Landau problem), and separately, the bound states of a particle in the background of a finite cylindrical well.

The article is organized as follows: in section \ref{sec:algebraicstructure} we introduce the non-commutative dynamical variables and define the generator of rotations in the non-commutative phase space. We then establish the conventions regarding the Dirac Hamiltonians in two dimensions we will consider throughout the article. In section \ref{sec:algebraicstructure2} we present a detailed construction of the generators of the graded Lie algebra $\mathfrak{sl}(2|1)$ in terms of the non-commutative phase-space variables, and the Dirac matrices in $2+1$ dimensions. Furthermore, we construct the total angular momentum operator which will generate an additional one dimensional Lie algebra $\mathfrak{so}(2)$ required to fully express the Hamiltonians under consideration. In section \ref{sec:irreps} we provide a complete characterization of the unitary irreducible representations of $\mathfrak{sl}(2|1)\oplus \mathfrak{so}(2)$ which we then employ in sections \ref{sec:Landau} and \ref{sec:cylinder} to solve the problems of a particle in a uniform magnetic field and in a finite cylindrical well respectively. Finally, three appendices are included to clarify some intermediary calculations.

\section{The Non-Commutative Two-Dimensional Phase Space}

\label{sec:algebraicstructure}
%We restrict to the phase $\kappa\theta < \hbar^2$ for it is the one which can be connected with the ordinary (commutative) case.

Non-commutativity on a two-dimensional phase space can be implemented by the following commutation relations between the Hermitian operators $\hat{X}_i$ and $\hat{P}_i$
\begin{equation}
\label{ec:relacionesnoconm}
    \left[\hat{X}_i,\hat{X}_j\right]=i \theta \epsilon_{ij}\,, \quad   \left[\hat{X}_i,\hat{P}_j\right]=i \delta_{ij}\,,\quad   \left[\hat{P}_i,\hat{P}_j\right]=i \kappa \epsilon_{ij} \,, \quad i=1,2
\end{equation}
where $\theta$ and $\kappa$ are the (real) non-commutativity parameters (we choose a unit system in which $\hbar=c=1$). With no loss of generality, we can take $\theta\geq 0$. These relations reduce to the Heisenberg algebra for ordinary dynamical variables in the $\theta,\kappa \to 0$ limit,
\begin{equation}
    \left[\hat{x}_i,\hat{x}_j\right]=0\,, \quad   \left[\hat{x}_i,\hat{p}_j\right]=i \delta_{ij}\,,\quad   \left[\hat{p}_i,\hat{p}_j\right]=0 \,.
\end{equation}
From equation \eqref{ec:relacionesnoconm}, it follows that the generator of rotations on the NC phase-space \cite{Nair} is given by 
\begin{equation}
\label{ec:momentoangularNC}
    \hat{L}:=\frac{1}{\left(1-{\kappa\theta}\right)}\left\{\left( \hat{X}_1 \hat{P}_2 - \hat{X}_2 \hat{P}_1\right)+
   {{  \frac{\theta}{2  } \left({\hat{P}_1}^2+{\hat{P}_2}^2\right)}+{ \frac{\kappa}{2  }
   \left({\hat{X}_1}^2+{\hat{X}_2}^2\right)}}\right\}\,,
\end{equation}
for $\kappa\theta \neq 1$. Indeed, it can be seen that $\hat{\mathbf{X}}$ and $\hat{\mathbf{P}}$ transform as vectors under rotations on the plane

\begin{equation}\label{PQvec}
     \left[\hat{L},\hat{X}_i\right] =i \, \epsilon_{i j} \hat{X}_j\,, \quad
     \left[\hat{L},\hat{P}_i\right] =i \, \epsilon_{i j} \hat{P}_j \,.
\end{equation}
Thus,

\begin{equation}\label{PQ2}
     \left[\hat{L},\hat{\mathbf{X}}^2\right] =0\,, \quad
     \left[\hat{L},\hat{\mathbf{P}}^2\right] =0\,.
\end{equation}
Furthermore, for the critical value $\kappa_c = 1/\theta$ (where $\hat{L}$ in equation \eqref{ec:momentoangularNC} is ill-defined), the commutator algebra in equation \eqref{ec:relacionesnoconm} reduces to
\begin{equation}
\label{ec:conmutconclu}
        \left[\hat{X}_i,\hat{X}_j\right]=i\epsilon_{ij}\theta \,, \quad \left[\hat{X}_i,\hat{P}_j\right]=i\delta_{ij}\, , \quad \left[\hat{P}_i,\hat{P}_j\right]=i\epsilon_{ij}\theta^{-1}\, ,
\end{equation}
which can be satisfied by just one pair of dynamical variables $\hat{X}_1=-\theta \hat{P}_2$ and $\hat{P}_1=\frac{\hat{X}_2}{\theta}$. Therefore, as a consequence of the degeneracy of the symplectic form in classical phase space for this particular value of $\kappa$, there occurs a \emph{dimensional reduction} \cite{Nair, Bellucci}. Indeed, two elements in the deformed Heisenberg algebra defined by equation \eqref{ec:relacionesnoconm},
\begin{equation}\label{magnetic_generators}
    \hat{P}=\hat{P}_2+\kappa \hat{X}_1, \quad \hat{Q}=\hat{X}_2-\theta \hat{P}_1,
\end{equation}
which are proportional to the \emph{magnetic translation generators} (See \cite{Falomir1}, equations (2.8-9), for example), commute with $\hat{X}_1$ and $\hat{P}_1$ and satisfy
\begin{equation}\label{QconP}
    \left[\hat{Q},\hat{P}\right]=i(1-\kappa\theta) .
\end{equation}
Thus, they are well-defined and belong to the center of the algebra when $\kappa = \kappa_c$, where the system becomes effectively one-dimensional.

The free Hamiltonian we will consider is of the standard Dirac-type, which in terms of the non-commutative momenta, may be expressed as
\begin{equation}
    \hat{H}=\boldsymbol{\alpha}\cdot \hat{\mathbf{P}}-\beta m
\end{equation}
where $\beta = \gamma^{0}$ and $\alpha_{i}=\beta \gamma^i$ with $\gamma^\mu$, $(\mu=0,1,2$) the Dirac matrices in $2+1$ space-time dimensions. These satisfy the Clifford algebra\footnote{We choose the signature $\eta^{\mu \nu}=\rm{diag}(1,-1,-1)$.} $\{\gamma^\mu,\gamma^\nu\}=2\eta^{\mu\nu}$  which can be realised by choosing
\begin{equation}
    \gamma^{0}=\sigma_3, \quad \gamma^{1}=i\sigma_2,  \quad \gamma^{2}=- i\sigma_1 \, .
\end{equation}
Thus, we have
\begin{equation}
\label{ec:hamiltoniano2d}
    \hat{H}=\hat{\mathbf{P}}\cdot \boldsymbol{\sigma}-m\sigma_3.
\end{equation}
which is the two dimensional non-commutative Dirac Hamiltonian of a free massive particle. Later on, we will also consider the Hamiltonian of the non-commutative version of an electron subject to a uniform magnetic field, and of an electron confined within a finite cylindrical well. 

%In the next section we develop an algebraic method to systematically study this type of Hamiltonians, with aggregated interactions, from an algebraic perspective. This procedure extends the one constructed in \cite{Falomir1} for rotationally invariant \textit{non-relativistic} Hamiltonians in the non-commutative two dimensional phase space. 

\section{Algebraic Structure of Dirac Hamiltonians}

\label{sec:algebraicstructure2}

In the previous section, we noticed that the non-singularity of the symplectic form establishes a critical value $\kappa_c=1/\theta$, that defines two different regions, namely $\kappa < \kappa_c$ and $\kappa> \kappa_c$. However, only the first one can be connected with the ordinary (commutative) case, and in the following we shall therefore restrict ourselves to this region of parameters. Bearing this in mind, we show that the characteristic subspaces of the Hamiltonians we will consider are characterized by unitary irreducible representations of the graded Lie algebra $\mathfrak{sl}(2|1)\oplus\mathfrak{so}(2)$.

%As a matter of fact, this is true regardless of whether $\kappa>0$ or $\kappa<0$. However, we will study these two cases separately. 

%\subsection{The $0<\kappa<\kappa_c$ case}
%\label{sec:hola3}

Taking $0<\kappa<\kappa_c$\footnote{As a matter of fact, $\kappa<0$ is a perfectly valid choice. However, the corresponding algebraic construction is very similar (\emph{cf.} \cite{Falomir1}) and we will therefore omit it.} we may define the generators $\mathcal{J}_0, \mathcal{J}_\pm$ as
\begin{equation}\label{algeb1}
    \begin{array}{c} \displaystyle
      \mathcal{J}_0= \frac{\sqrt{{\kappa\theta }}}{4} 
   \left(\frac{\mathbf{{{\hat{P}}}}^2}{\kappa
   }+\frac{\mathbf{{{\hat{X}}}}^2}{\theta }+
   {2 \hat{L}}\right) = {\mathcal{J}_0}^\dagger\,,
    \\ \\ \displaystyle
      \mathcal{J}_\pm=\frac{\sqrt{{\kappa\theta }}}{4 \sqrt{1-{\theta  \kappa }}} \left\{\frac{\mathbf{{{\hat{X}}}}^2}{\theta }
      -\frac{\mathbf{{{\hat{P}}}}^2}{\kappa }  \mp
   \frac{i }{\sqrt{\kappa\theta
   }}\left(\mathbf{\hat{P}}\cdot \mathbf{\hat{X}}+\mathbf{\hat{X}}\cdot \mathbf{\hat{P}}\right) \right\}=
   {\mathcal{J}_\mp}^\dagger \,,
    \end{array}
\end{equation}
that satisfy the commutator algebra of $\mathfrak{sl}(2,\mathbb{R})$

\begin{equation}\label{algeb2}
    \left[\mathcal{J}_0 , \mathcal{J}_\pm\right] = \pm \mathcal{J}_\pm\,, \quad
    \left[\mathcal{J}_+ , \mathcal{J}_-\right] = -2\mathcal{J}_0 \,,
\end{equation}
where $\mathcal{J}_\pm=\mathcal{J}_1\pm i \mathcal{J}_2$, with $\mathcal{J}_i, \, \, (i=0,1,2)$, Hermitian operators. Moreover, the quadratic Casimir operator  ${\mathcal{J}}^2=\mathcal{J}_0(\mathcal{J}_0\mp 1)-\mathcal{J}_\pm \mathcal{J}_\mp=\mathcal{J}_0^2-\mathcal{J}_1^2-\mathcal{J}_2^2$ is given by

\begin{equation}\label{algeb3}
   \begin{array}{c} \displaystyle
      {\mathcal{J}}^2=\frac{{\kappa\theta}}{16\left( 1- {\kappa\theta} \right)}
    \left\{\left( 1-{\kappa\theta}\right)
    \left( \frac{\mathbf{\hat{X}}^2}{\theta} +  \frac{\mathbf{\hat{P}}^2}{\kappa} + 2 {\hat{L}}\right)^2  - \right.
    \\ \\ \displaystyle
     \left. -  \left( \frac{\mathbf{\hat{X}}^2}{\theta} -  \frac{\mathbf{\hat{P}}^2}{\kappa} \right)^2 -
     \frac{\left( \mathbf{\hat{P}}\cdot \mathbf{\hat{X}}+\mathbf{\hat{X}}\cdot \mathbf{\hat{P}} \right)^2}{\theta \kappa} \right\} \,,
   \end{array}
\end{equation}
which notably, is not a positive definite operator \cite{Bargmann}. Furthermore, equations \eqref{PQvec}-\eqref{PQ2} imply

\begin{equation}\label{algeb5}
    \left[ \hat{L} , \mathcal{J}_0 \right]=0\,, \quad \left[ \hat{L} , \mathcal{J}_\pm \right]=0
    \quad \Longrightarrow \quad \left[ \hat{L} , {\mathcal{J}}^2 \right]=0 \,.
\end{equation}
In fact, one can also verify that \cite{Falomir1}

\begin{equation}\label{JcuadLcuad}
    \mathcal{J}^2=\frac{1}{4}\left\{ \hat{L}^2-1 \right\} \geq - \frac{1}{4}\,,
\end{equation}
which imposes a constraint on the irreducible representations of the Lie algebra $\mathfrak{sl}(2,\mathbb{R})$. In particular, this relation implies that only the \textit{discrete classes} of irreducible unitary representations of $\mathfrak{sl}(2,\mathbb{R})$ will be of interest for our purposes \cite{Bargmann, Gamboa2}. The  construction we have described above has been done for the bosonic case in \cite{Falomir1}.

%In section \ref{sec:cilindro} we'll consider the problem of a cylindrical potential well, and thus, it is relevant to express:

%\begin{equation}\label{ec:equiscuadrado}
 %   \frac{\hat{\mathbf{X}}^2}{\theta}=2\sqrt{\frac{1}{\kappa\theta}}\mathcal{J}_0+\sqrt{\frac{1}{\kappa\theta}-1}\left(\mathcal{J}_++\mathcal{J}_-\right)-\hat{L}
%\end{equation}
From equation \eqref{ec:hamiltoniano2d}, we note that the Dirac Hamiltonian in $2+1$ dimensions contains a kinetic term, $\hat{\mathbf{P}}\cdot \boldsymbol{\sigma}$, which is linear in momentum. Furthermore, in the next sections, we will discuss problems with operators of the form $\hat{\mathbf{X}}\times \boldsymbol{\sigma} = \epsilon^{ij}\hat{X}_i \sigma_j$, that is, linear in the NC coordinate operators. In analogy to what is usually done in supersymmetric theories, we consider the $\mathbb{Z}_2-$graded extension of $\mathfrak{sl}(2,\mathbb{R})$ in order to incorporate fermionic-type operators. We start by defining
\begin{equation}\label{ec:definicionsigma}
    \Sigma:=\hat{L}+\sigma_3
\end{equation}
which, together with $\mathcal{J}^2$ and $\mathcal{J}_\pm$ generate the \emph{even} subspace $\mathfrak{u}(1)\oplus \mathfrak{sl}(2,\mathbb{R})$ of the graded Lie algebra. To construct the \emph{odd} generators, we define first

\begin{equation}
\label{ec:defq}
\begin{array}{c}
\mathcal{Q}_{\pm}:=\displaystyle{\frac{1}{2}\sqrt[4]{ \frac{\kappa\theta}{1-\kappa\theta}}}
    \left\{\left( \frac{\hat{X}_1}{\sqrt{\theta}} \mp i \, \frac{\hat{P}_1}{\sqrt{\kappa}}\right)\sigma_1 + \left(\frac{\hat{X}_2}{\sqrt{\theta}} \mp i \, \frac{\hat{P}_2}{\sqrt{\kappa}}\right)\sigma_{2}\right\}\,, \\
    \, \\
    \Tilde{\mathcal{Q}}_{\pm}:=\sigma_3 \mathcal{Q}_{\pm} \,.
\end{array}
\end{equation}
In terms of $\mathcal{Q}_\pm$ and $\Tilde{\mathcal{Q}}_\pm$ we define a pair of fermionic multiplets $(\mathcal{R}_+,\mathcal{R}_-)$ and $(\mathcal{S}_+,\mathcal{S}_-)$ given by

\begin{equation}\label{def-RyS}
\begin{split}
   & \mathcal{R}_+:=\frac12 \sqrt[4]{\frac{1- \sqrt{\kappa\theta}}{1+ \sqrt{\kappa\theta}}} \left(  \mathcal{Q}_+ + \widetilde{\mathcal{Q}}_+\right) \,,  \quad
     \mathcal{R}_-:=\frac12 \sqrt[4]{\frac{1+ \sqrt{\kappa\theta}}{1 - \sqrt{\kappa\theta}}} \left(  \mathcal{Q}_- + \widetilde{\mathcal{Q}}_-\right)  \\
    &  \mathcal{S}_+:=\frac12 \sqrt[4]{\frac{1+ \sqrt{\kappa\theta}}{1- \sqrt{\kappa\theta}}} \left(  \mathcal{Q}_+ - \widetilde{\mathcal{Q}}_+\right) \,,  \quad
     \mathcal{S}_-:=\frac12 \sqrt[4]{\frac{1- \sqrt{\kappa\theta}}{1 + \sqrt{\kappa\theta}}} \left(  \mathcal{Q}_- - \widetilde{\mathcal{Q}}_-\right) \,.
     \end{split}
\end{equation}
As expected, the fermionic elements are nilpotent,

\begin{equation}\label{ec:nilpotencia}
    \mathcal{R}_\pm^2=\mathcal{S}_\pm^2=0 \,, 
\end{equation}
and they satisfy

\begin{equation}\label{ec:prodcruz}
    \mathcal{R}_+\mathcal{R}_-=\mathcal{S}_+\mathcal{S}_-=0\,, \quad \mathcal{R}_+^\dagger=\mathcal{S}_-\,, \quad\mathcal{R}_-^\dagger=\mathcal{S}_+ \,.
\end{equation}
Thus, the set of eight generators $\left\{\mathcal{J}_0, \mathcal{J}_{\pm}, \Sigma, \mathcal{R}_\pm, \mathcal{S}_\pm\right\}$ span the graded Lie algebra $\mathfrak{sl}(2|1)$, whose (anti)commutation relations are given by\footnote{We have chosen different conventions than, for example, in \cite{Schomerus}.}
\begin{equation}
\label{ec:NCcomutationrelations}
    \begin{array}{c}
         \left[ \Sigma ,{\mathcal{R}}_\pm  \right] = {\mathcal{R}}_\pm \,, \quad \quad  \left[ \Sigma ,{\mathcal{S}}_\pm  \right] = -{\mathcal{S}}_\pm  \,,     \\ \\
         \left[  \mathcal{J}_0, {\mathcal{R}}_\pm  \right] =  \pm \frac 12 {\mathcal{R}}_\pm \,, \quad \quad \left[  \mathcal{J}_0, {{\mathcal{S}}}_\pm  \right] = \pm \frac 12 {{\mathcal{S}}}_\pm \,,
         \\ \\
         \left[ \mathcal{J}_\pm , {\mathcal{R}}_\mp  \right] = \mp {\mathcal{R}}_\pm \,, \quad \quad
         \left[ \mathcal{J}_\pm , {\mathcal{S}}_\mp  \right] = \mp {\mathcal{S}}_\pm \,, \\ \\
          \left[ \mathcal{J}_0 , \mathcal{J}_\pm \right] = \pm \mathcal{J}_\pm \,, \quad\quad \left[ \mathcal{J}_+ , \mathcal{J}_- \right]= -2 \mathcal{J}_0 \,,\\ \\
         \left\{ {\mathcal{R}}_\pm ,  {\mathcal{S}}_\pm \right\} =\mathcal{J}_\pm \,, \quad \quad  \left\{ {\mathcal{R}}_\pm ,  {\mathcal{S}}_\mp \right\} =\mathcal{J}_0\mp \frac{\Sigma}{2} \,, \\
    \end{array}
\end{equation}
where unspecified (anti)commutators vanish. Next, we define a \emph{total} angular momentum operator 
\begin{equation}\label{ec:definicionele}
    \mathcal{L}:=\hat{L}+\frac{\sigma_3}{2}\,
\end{equation}
which commutes with all the generators of $\mathfrak{sl}(2|1)$ and that will be required to express the Hamiltonians we will consider in what follows. As a matter of fact, these will be completely expressible in terms of $\hat{\mathbf{P}}\cdot \boldsymbol{\sigma}$, $\hat{\mathbf{X}}\times \boldsymbol{\sigma}$, $\hat{\mathbf{X}}^2$, $\Sigma$ and $\mathcal{L}$ itself. We thus write for future use, 

\begin{equation}\label{VD-2}
    \begin{array}{c}\displaystyle
      \hat{\mathbf{P}}\cdot \boldsymbol{\sigma}=\hat{P}_i \, \sigma_i =  i \sqrt[4]{\frac{\kappa }{\theta}} \left[\sqrt{1+\sqrt{\kappa\theta}}
   \left( \mathcal{R}_+ -\mathcal{S}_- \right)-\sqrt{1-\sqrt{\kappa\theta} }
   \left( \mathcal{R}_- -\mathcal{S}_+ \right)\right]\, ,
      \\ \\ \displaystyle
      \hat{\mathbf{X}}\times \boldsymbol{\sigma}=\epsilon^{ij}\hat{X}_i \, \sigma_j=
      -i \sqrt[4]{\frac{\theta }{\kappa}} \left[\sqrt{1+\sqrt{\kappa\theta}}
   \left( \mathcal{R}_+ - \mathcal{S}_- \right)+\sqrt{1-\sqrt{\kappa\theta} }
   \left( \mathcal{R}_- -  \mathcal{S}_+ \right)\right] \,.
    \end{array} 
\end{equation}
In this way, each characteristic subspace of the Hamiltonians is a representation space of a unitary irreducible representation of the direct sum of Lie algebras $\mathfrak{sl}(2|1)\oplus \mathfrak{so}(2)$, where the second factor is generated by $\mathcal{L}$.

\section{The Irreducible Representations of $\mathfrak{sl}(2|1)\oplus \mathfrak{so}(2)$}
\label{sec:irreps}

In this section we present a detailed construction of the irreducible representations of the direct sum of Lie algebras $\mathfrak{sl}(2|1)\oplus \mathfrak{so}(2)$. We start with the unitary irreducible representations in the discrete classes of the Lie algebra $\mathfrak{sl}(2,\mathbb{R})$, restriction due to the constraint on $\mathcal{J}^2$ being bounded below (\emph{cf.} equation \eqref{JcuadLcuad} and \cite{Gamboa2, Bargmann}). These representations can be characterized by the eigenvalue of the Casimir operator $\mathcal{J}^2$ which takes values $\lambda=k(k-1)$ with $k=\tfrac{1}{2},1,\tfrac{3}{2},\dots$ . For a given $k$ the representation space is generated by the simultaneous eigenvectors of $\mathcal{J}^2$ and $\mathcal{J}_0$, 

\begin{equation}
   \begin{array}{c}
        \mathcal{J}^2|k,m\rangle=k(k-1)|k,m\rangle   \\
       \, \\
        \mathcal{J}_0|k,m\rangle=m|k,m\rangle \,,\quad \text{with}\quad m=k,k+1,\dots \text{ or } m=-k,-(k+1),\dots \,.\\
   \end{array}
\end{equation}
We recall that the even subspace of the graded extension has an additional $\mathfrak{u}(1)$ factor generated by $\Sigma$ and thus we consider the vectors which are also eigenvectors of this operator. We denote these by $|k,m,\sigma\rangle$, such that

\begin{equation}
   \begin{array}{c}
        \Sigma|k,m,\sigma\rangle=\sigma|k,m,\sigma\rangle  \,,\quad \text{with }\sigma \in \mathbb{R} \,. \\
   \end{array}
\end{equation}
We denote $\langle k,\sigma\rangle={\rm span}(\{|k,m,\sigma\rangle\})$.

Next we characterize the irreducible representation spaces of the graded Lie algebra, which are the direct sum of a finite number of irreducible representation spaces of the even subalgebra,
\begin{equation}
    \langle k_0,\sigma_0 \rangle \oplus \langle k_1,\sigma_1\rangle \oplus \dots \oplus \langle k_n,\sigma_n \rangle \quad \text{with} \quad k_0 \leq k_1 \leq \dots \leq k_n \,.
\end{equation}
To determine which values of $k$ appear in this direct sum decomposition we use the fermionic generators. At this point, it is worth stressing that the vectors $|k_i,m_i,\sigma_i\rangle \in \langle k_i,\sigma_i\rangle$ will have values of $m_i = k_i,k_i+1,\dots$ or $m_i=-k_i,-(k_i+1),\dots$ but not both. In particular, without loss of generality, the construction we present assumes that $m_0=k_0,k_0+1,\dots$ , given that the negative $m_0$ case follows in exactly the same way. Notably, in \S \ref{sec:so2factor}, we will argue that for the physical systems we are considering, we require the value of $m$ to be bounded below and thus, in practice, we will only make use of the representations we construct explicitly in this section.

So let us consider the \emph{minimum} weight vector $|k_0,k_0,\sigma_0\rangle \in \langle k_0,\sigma_0\rangle$. Using the commutation relations \eqref{ec:NCcomutationrelations}, it is straightforward to see that

\begin{equation}
\begin{array}{c}
    \mathcal{J}_0\mathcal{R}_-|k_{0},k_{0},\sigma_0 \rangle = \left(k_{0}-\tfrac{1}{2}\right)\mathcal{R}_-|k_{0},k_{0},\sigma_0 \rangle\,,
\end{array}
\end{equation}
from where we conclude that 
%This would imply that $\mathcal{R}_-|k_{0},k_{0},\sigma \rangle$ belongs to a representation subspace with $k<k_{0}$, in contradiction with $\langle k_{0}, \sigma\rangle$ being the subspace with the minimum $k_{0}$ among the other (representation) subspaces in the graded representation space. We thus take 
$\mathcal{R}_-|k_{0},k_{0},\sigma_0\rangle=0$. With a similar argument, we obtain $\mathcal{S}_-|k_{0},k_{0},\sigma_0\rangle=0$. We now proceed with $\mathcal{R}_+$. By considering the action of the product of operators $\mathcal{J}_0\mathcal{R}_+$, $\Sigma\mathcal{R}_+$ and using the commutation relations \eqref{ec:NCcomutationrelations} it is easy to verify that

\begin{equation}
    \mathcal{R}_+|k_{0},k_{0},\sigma_0\rangle=\sqrt{k_{0}-\tfrac{\sigma_0}{2}}|k_{0}+\tfrac{1}{2},k_{0}+\tfrac{1}{2},\sigma_0+1\rangle \in \langle k_{0}+\tfrac{1}{2},\sigma_0+1 \rangle \,,
\end{equation}
which is nonvanishing as long as $\sigma_0 \neq 2k_0$. Moreover, by successively applying $\mathcal{J}_+$, we construct a complete basis for the subspace $\langle k_{0}+\tfrac{1}{2},\sigma_0+1 \rangle$. With similar arguments we have

\begin{equation}
    \mathcal{S}_+|k_{0},k_{0},\sigma_0\rangle=\sqrt{k_{0}+\tfrac{\sigma_0}{2}}|k_{0}+\tfrac{1}{2},k_{0}+\tfrac{1}{2},\sigma_0-1\rangle  \in \langle k_{0}+\tfrac{1}{2},\sigma_0-1 \rangle \,,
\end{equation}
which is nonvanishing as long as $\sigma_0 \neq -2k_0$. Again, with the successive application of $\mathcal{J}_+$ we construct the complete basis for the subspace $\langle k_{0}+\tfrac{1}{2},\sigma_0-1 \rangle$. We further recall that the odd generators are nilpotent and that their crossed products are identically vanishing (equations \eqref{ec:nilpotencia} and \eqref{ec:prodcruz}) and thus

\begin{equation}
    \mathcal{R}_\pm|k_{0}+\tfrac{1}{2},k_{0}+\tfrac{1}{2},\sigma_0+1\rangle=0=\mathcal{S}_\pm|k_{0}+\tfrac{1}{2},k_{0}+\tfrac{1}{2},\sigma_0-1\rangle\,.
\end{equation}
But, from equation  \eqref{ec:NCcomutationrelations}, we have

\begin{equation}
    \begin{array}{c}
     \mathcal{R}_-|k_{0}+\tfrac{1}{2},k_{0}+\tfrac{1}{2},\sigma_0-1\rangle = \sqrt{k_{0}+\tfrac{\sigma_0}{2}}|k_{0},k_{0},\sigma_0\rangle  \\
     \\
     \mathcal{S}_-|k_{0}+\tfrac{1}{2},k_{0}+\tfrac{1}{2},\sigma_0+1\rangle = \sqrt{k_{0}-\tfrac{\sigma_0}{2}}|k_{0},k_{0},\sigma_0\rangle 
    \end{array}
\end{equation}
and therefore, the action of $\mathcal{S}_-$ or $\mathcal{R}_-$ on vectors in $\langle k_{0}+\tfrac{1}{2},\sigma_0+1 \rangle$ and in $\langle k_{0}+\tfrac{1}{2},\sigma_0-1\rangle$ respectively, takes us back to $\langle k_{0},\sigma_0\rangle$. 

Finally, we are left with the analysis of $\mathcal{R}_+\mathcal{S}_+|k_{0},k_{0},\sigma\rangle$ and $\mathcal{S}_+\mathcal{R}_+|k_{0},k_{0},\sigma\rangle$. In contrast with the previous cases, these vectors are not orthogonal to the subspace $\langle k_{0},\sigma_0\rangle$; for instance, we can take the linear combination

\begin{equation}
\label{ec:holaJcero}
    \{\mathcal{R}_+,\mathcal{S}_+\}|k_{0},k_{0},\sigma_0\rangle=\mathcal{J}_+|k_{0},k_{0},\sigma_0\rangle=\sqrt{2k_{0}}|k_{0},k_{0}+1,\sigma_0\rangle \in \langle k_{0},\sigma_0\rangle \,.
\end{equation}
Indeed, to construct a vector orthogonal to the minimal weight subspace we look for the value of $a$ such that

\begin{equation}
\begin{split}
      0&=\langle k_0,k_0+1,\sigma_0|\left( \mathcal{S}_+\mathcal{R}_+|k_0,k_0,\sigma_0\rangle -a|k_0,k_0+1,\sigma_0\rangle \right)\\
      &=\frac{1}{\sqrt{2k_0}}\langle k_0,k_0,\sigma_0|\mathcal{J}_+^\dagger \mathcal{S}_+\mathcal{R}_+|k_0,k_0,\sigma_0\rangle - a\\
      &=\frac{1}{\sqrt{2k_0}}\|\mathcal{R}_+|k_0,k_0,\sigma_0\rangle\|^2 - a\\
      &=\frac{1}{\sqrt{2k_0}}(k_0-\tfrac{\sigma_0}{2})-a\,,
\end{split}
\end{equation}
where we have used \eqref{ec:holaJcero} and \eqref{ec:NCcomutationrelations}. Then, $a=\frac{1}{\sqrt{2k_0}}\left(k_0-\tfrac{\sigma_0}{2}\right)$ is the adequate choice. Moreover, by considering the action of the operators $\mathcal{J}_0$ and $\Sigma$ on this vector, and taking its norm, we conclude that
\begin{equation}
\label{ec:finalpro}
\begin{split}
\left(\mathcal{S}_+\mathcal{R}_+-\tfrac{k_0-\sigma_0/2}{2k_0}\mathcal{J}_+\right)&|k_0,k_0,\sigma_0\rangle\\
&
    =\sqrt{\tfrac{(2k_0+1)}{8k_0}\left(4k_0^2-\sigma_0^2\right)}|k_0+1,k_0+1,\sigma_0\rangle \in \langle k_0+1,\sigma_0\rangle\,.    
\end{split}
\end{equation}
Notice that if $\sigma_0 =\pm 2k_0$, then $\left(\mathcal{S}_+\mathcal{R}_+-\frac{k_0-\sigma_0/2}{2k_0}\mathcal{J}_+\right)|k_0,k_0,\sigma_0\rangle=0$. Once again, with the successive application of $\mathcal{J}_+$ we construct a complete basis for the subspace $\langle k_0+1,\sigma_0\rangle$. With equation \eqref{ec:finalpro} we have exhausted all of the possibilities, given that any additional combination of products of fermionic operators will vanish, due to equations \eqref{ec:nilpotencia} and \eqref{ec:prodcruz}.

All in all, we observe that the irreducible representations of the graded Lie algebra $\mathfrak{sl}(2|1)$ may be characterized by $k=\tfrac{1}{2},1,\tfrac{3}{2},\dots$, the integer or half-integer appearing in the expression of the eigenvalue of the quadratic Casimir operator of the even subalgebra $\mathcal{J}^2$. For a given value of $k$, if $\sigma\neq \pm 2k$, the representation space is the direct sum of four different representation spaces of $\mathfrak{u}(1)\oplus\mathfrak{sl}(2,\mathbb{R})$

\begin{equation}
    \langle k, \sigma\rangle \oplus \langle k+\tfrac{1}{2},\sigma+1\rangle \oplus \langle k+\tfrac{1}{2},\sigma-1\rangle\oplus \langle k+1,\sigma\rangle\,,
\end{equation}
whereas if $\sigma=\pm 2k$, the graded representation space is the direct sum of only two of these representation spaces

\begin{equation}
\begin{array}{c}
     \langle k, \sigma\rangle \oplus \langle k+\tfrac{1}{2},\sigma-1\rangle\, \quad \quad \text{if} \quad \quad \sigma=2k \\ \\
     \langle k, \sigma\rangle \oplus \langle k+\tfrac{1}{2},\sigma+1\rangle\, \quad \quad \text{if} \quad \quad \sigma=-2k
\end{array}
\end{equation}

In appendix \ref{sec:apendiceaccoddgen} we compute the action of the fermionic generators in a general vector of each representation space for $\sigma=\pm 2k$.  

\subsection{Quadratic Casimir Operator of $\mathfrak{sl}(2|1)$}
\label{sec:casimiroperator}

Having constructed the irreducible representations of $\mathfrak{sl}(2|1)$, we compute the quadratic Casimir operator $\mathcal{C}^2$ of the graded Lie algebra. This can be done using the \emph{Killing form} $\mathbf{K}$ of the graded Lie algebra, whose components are \cite{Cornwell}

\begin{equation}
    K_{ab}:= {\rm Str}[T_aT_b]={\rm Tr}[T_aT_b   \mathcal{I}]
\end{equation}
where $T_a,T_b$ are the generators of $\mathfrak{sl}(2|1)$ in the adjoint representation, with $a,b=1,\dots 4$ corresponding to the even subalgebra, $a,b=5,\dots,8$ to the odd generators, and the matrix $\mathcal{I}$ is defined as

\begin{equation}
    \mathcal{I}:={\rm diag}(1,1,1,1,-1,-1,-1,-1)\,.
\end{equation}
In Appendix \ref{sec:apendiceadjunt} we write down the generators in the adjoint representation as well as the Killing form. This, in turn, can be used to calculate the quadratic Casimir operator as
\begin{equation}
\begin{split}
    \mathcal{C}^{2}
    &=\left(\mathcal{J}_0,\mathcal{J}_+,\mathcal{J}_-,\Sigma, \mathcal{R}_+,\mathcal{R}_-,\mathcal{S}_+,\mathcal{S}_-\right)\mathbf{K}^{-1}\left(\mathcal{J}_0,\mathcal{J}_+,\mathcal{J}_-,\Sigma, \mathcal{R}_+,\mathcal{R}_-,\mathcal{S}_+,\mathcal{S}_-\right)^{t}\\
    &=\mathcal{J}_0^2-\tfrac{1}{2}\left(\mathcal{J}_+\mathcal{J}_-+\mathcal{J}_-\mathcal{J}_+\right)+\mathcal{J}_0-\left(\mathcal{R}_+\mathcal{S}_-+\mathcal{S}_+\mathcal{R}_-\right)-\tfrac{1}{4}\Sigma^2\,.    
\end{split}
\end{equation}
Now, if we act with $\mathcal{C}^2$ on the minimum weight vector\footnote{Note that given that $\mathcal{C}^2$ commutes with every generator of $\mathfrak{sl}(2|1)$, the eigenvalue of the minimum weight vector of $\langle k,\sigma \rangle$ characterizes the whole representation space.} of the representation space $\langle k,\sigma\rangle$ we get (\emph{cf.} appendix \ref{sec:apendiceaccoddgen} for the action of the odd generators)
\begin{equation}
\label{ec:chau1}
   \mathcal{C}^2|k,k,\sigma\rangle 
   =\left(k^2-\frac{\sigma^2}{4}\right)|k,k,\sigma\rangle \,.
\end{equation}
On the other hand, using the commutation relations in \eqref{ec:NCcomutationrelations} and the defining relations of the generators (equation \eqref{def-RyS}), it is straightforward to verify that, in terms of the dynamical variables in \eqref{ec:relacionesnoconm}, the quadratic Casimir operator can be written as
\begin{equation}
\label{ec:chau2}
    \mathcal{C}^{2}=\mathcal{J}^2-\tfrac{1}{4}\left(\hat{L}^2-1\right) =0 \,,
\end{equation}
where the last equality follows from equation \eqref{JcuadLcuad}. Thus, from equations \eqref{ec:chau1} and \eqref{ec:chau2} we conclude that the allowed values of $\sigma$ are precisely $\sigma=\pm 2k$. This implies that, taking into account our previous construction of the representation space of $\mathfrak{sl}(2|1)$, the representation space of the \emph{irreps} we will be interested in are restricted to

\begin{equation}
\label{ec:espaciocasos1}
\begin{cases}
     \langle k, \sigma\rangle \oplus \langle k+\tfrac{1}{2},\sigma-1\rangle \quad \text{if} \quad \sigma = 2k \\
     \\
       \langle k, \sigma\rangle \oplus \langle k+\tfrac{1}{2},\sigma+1\rangle \quad \text{if} \quad \sigma = -2k \,.
\end{cases}
\end{equation}

\subsection{The $\mathfrak{so}(2)$ factor}
\label{sec:so2factor}

As we discussed in \S \ref{sec:algebraicstructure2}, in order to express the Hamiltonians we will consider in this article, we require an additional operator $\mathcal{L}$. This operator commutes with all of the generators of $\mathfrak{sl}(2|1)$ so that we take it to be the generator of an additional one dimensional Lie algebra $\mathfrak{so}(2)$. The irreducible representations of this Lie algebra are characterized by a real number\footnote{Below we show that in our setting, $j$ will actually take half-integer values.} $j\in \mathbb{R}$ such that, for a given $j$, the representation space is the one dimensional vector space generated by $|j\rangle$ with $\mathcal{L}|j\rangle=j|j\rangle$. 

Thus, the characteristic subspaces of the Hamiltonians we will study are representation spaces of unitary irreducible representations of the direct sum of Lie algebras $\mathfrak{sl}(2|1)\oplus \mathfrak{so}(2)$, which we denote by $\langle k,\sigma, j\rangle$, whose vectors we write in the form $|k,m,\sigma,j\rangle$. In particular, taking into account the discussion of the representations selected by the quadratic Casimir operator in \S \ref{sec:casimiroperator}, we have just to consider
\begin{equation}
\label{ec:espaciocasos2}
\begin{cases}
     \langle k, \sigma, j\rangle \oplus \langle k+\tfrac{1}{2},\sigma-1,j\rangle \quad \text{if} \quad \sigma = 2k \\
     \\
       \langle k, \sigma, j\rangle \oplus \langle k+\tfrac{1}{2},\sigma+1,j\rangle \quad \text{if} \quad \sigma = -2k \,.
\end{cases}
\end{equation}
Notably, given that the vectors of the above representation spaces are eigenvectors of both $\Sigma$ and $\mathcal{L}$, they are also eigenvectors of the linear combinations $\Sigma_3:=2(\Sigma-\mathcal{L})$ and $L:=2\mathcal{L}-\Sigma$. In fact, for $\sigma=2k$ we have,
\begin{equation}
    \begin{array}{c}
         \Sigma_3 |k,m,2k,j\rangle=2(2k-j)|k,m,2k,j\rangle \\
         \\
         \Sigma_3 |k+\tfrac{1}{2},m,2k-1,j\rangle=2(2k-1-j)|k+\tfrac{1}{2},m,2k-1,j\rangle \,.\\
    \end{array}
\end{equation}
In particular, for the applications we are interested in, $\Sigma_3$ is represented by the third Pauli matrix $\sigma_3$ (\emph{cf.} equations \eqref{ec:definicionsigma} and \eqref{ec:definicionele}), whose eigenvalues are $\pm 1$. Thus, we conclude that in our setting, $2(2k-j)=1$ while $2(2k-1-j)=-1$, which gives us an expression for $j$ in terms of $k$, given by the half-integer $j=2k-\tfrac{1}{2}$. Similarly, we may act on the vectors of the representation space corresponding to $\sigma=2k$ with $L$, to get
\begin{equation}
    \begin{array}{c}
         L |k,m,2k,2k-\tfrac{1}{2}\rangle= (2k-1)|k,m,2k,2k-\tfrac{1}{2}\rangle \\
         \\
         L |k+\tfrac{1}{2},m,2k-1,2k-\tfrac{1}{2}\rangle=2k|k+\tfrac{1}{2},m,2k-1,2k-\tfrac{1}{2}\rangle \,.
    \end{array}
\end{equation}
In the representation defined by the dynamical variables \eqref{ec:momentoangularNC}, $L$ is represented by $\hat{L}$ (\emph{cf.} equations \eqref{ec:definicionsigma} and \eqref{ec:definicionele}), so the eigenvalue $\ell$ of $\hat{L}$ is related to the eigenvalue of $\mathcal{J}^2$ by $\ell=2k-1\geq 0$, consistent with equation \eqref{JcuadLcuad}. On the other hand, if $\sigma=-2k$, with a completely analogous calculation, we get $j=\tfrac{1}{2}-2k$ and $\ell=-(2k-1)< 0$, which again, is consistent with equation \eqref{JcuadLcuad}. 

Taking this into account, and in order to simplify notation for the examples we will consider in the next sections, we relabel our vectors using a different set of quantum numbers, namely, $k$, $\ell$ and $\pm$, the latter, corresponding to the eigenvalue of $\sigma_3$. Explicitly,
\begin{equation}
\label{ec:horacio1}
    |k,m,\sigma,j\rangle \longrightarrow |k,m,\ell,\pm \rangle\,.
\end{equation}
Note that in terms of these quantum numbers, $\sigma=2k$ and $\sigma=-2k$ correspond to positive and negative angular momentum respectively, so that equation \eqref{ec:espaciocasos2} becomes

\begin{equation}
\label{ec:horacio2}
    \begin{cases}
     \langle k, \ell, +\rangle \oplus \langle k+\tfrac{1}{2},\ell+1,-\rangle \quad \text{if} \quad \ell=2k-1\geq 0 \\
     \\
       \langle k, \ell, -\rangle \oplus \langle k+\tfrac{1}{2},\ell-1,+\rangle \quad \text{if} \quad \ell=-2k+1\leq 0 \,.
\end{cases}
\end{equation}

Having included the $\mathfrak{so}(2)$ factor, we may now argue why the eigenvalues of $\mathcal{J}_0$ have to be positive. From equation \eqref{algeb1} we may solve for 
\begin{equation}
    \frac{\hat{\mathbf{X}}^2}{\theta}=\frac{2}{\sqrt{\kappa \theta}}\mathcal{J}_0+\sqrt{\frac{1-\kappa \theta}{\kappa \theta}}(\mathcal{J}_++\mathcal{J}_-)-\hat{L} \,,
\end{equation}
and taking into account that this is the square of a Hermitian operator, we have
\begin{equation}
    0\leq \langle k,m,\ell,\pm|\hat{\mathbf{X}}^2|k,m,\ell,\pm \rangle 
    = \frac{2m}{\sqrt{\kappa \theta}}-\ell\,,
\end{equation}
from where we conclude that, for any given $\ell$ (positive or negative) $2m \geq \ell \sqrt{\kappa\theta}$. This implies that $m$ is bounded from below and thus $m=k,k+1,\dots$.

In the following sections we use this algebraic construction to solve two different problems: A charged particle in a uniform magnetic field, and a particle in a finite cylindrical well. 

\section{The Landau Problem}

\label{sec:Landau}

Let us consider the problem of a charged massive particle of spin $1/2$ in a uniform magnetic field perpendicular to the NC plane. We consider the Hamiltonian given in \eqref{ec:hamiltoniano2d} with a minimal coupling to the electromagnetic field,
\begin{equation}
    \hat{P_i}\to \hat{P}_i-q\hat A_i \,,
\end{equation}
where we choose
\begin{equation}
\label{ec:symmetricgauge}
    \hat A_i = -\frac{\mathcal B}{2}\epsilon_{ij}\hat{X}_j\,,
\end{equation}
for some real constant $\mathcal{B}$. Note that, according to the covariant definition of the field tensors, this gauge field produces a uniform magnetic field,
\begin{equation}
	B=\frac1{iq}\,[P_1-qA_1,P_2-qA_2]=\mathcal B +\frac\kappa{q}+\frac{\theta q\mathcal B^2}{4}\,.
\end{equation}
The Landau Hamiltonian reads
\begin{equation}\label{ec:hamiltonianolandau}
\hat{H}=\hat{\mathbf{P}}\cdot \boldsymbol{\sigma}-\omega \hat{\mathbf{X}}\times \boldsymbol{\sigma}-m\sigma_3 \,, \quad \text{with }\omega=\frac{q\mathcal B}{2}\,.
\end{equation}
For the moment we do not make any assumption on whether $q\mathcal B$ is positive or negative. 

%\subsection{The Landau Hamiltonian for $0<\kappa<\kappa_c$} In this region, 

Using Eq. \eqref{VD-2} we may express the Hamiltonian as
\begin{equation}
    \begin{array}{c}
      \hat{H}=\displaystyle i\sqrt[4]{\frac{\kappa}{\theta}}\left(\sqrt{1+\sqrt{\kappa\theta}}\left(\mathcal{R}_+-\mathcal{S}_-\right)-\sqrt{1-\sqrt{\kappa\theta}}\left(\mathcal{R}_--\mathcal{S}_+\right)\right)\\
    \displaystyle +i\omega\sqrt[4]{\frac{\theta}{\kappa}}\left(\sqrt{1+\sqrt{\kappa\theta}}\left(\mathcal{R}_+-\mathcal{S}_-\right)+\sqrt{1-\sqrt{\kappa\theta}}\left(\mathcal{R}_--\mathcal{S}_+\right)\right)-m\sigma_3  \,.    \\
    \end{array}
\end{equation}
We now apply a unitary transformation in order to simplify the expression to ultimately be able to diagonalize the Hamiltonian, 
\begin{equation}
\label{ec:transfunitaria}
\begin{split}
   \hat{H}_r
    &=e^{i\alpha \mathcal{J}_2}\hat{H}e^{-i\alpha \mathcal{J}_2}\\
    &=e^{i\alpha \mathcal{J}_2}\hat{H}_{m=0}e^{-i\alpha \mathcal{J}_2}-m\sigma_3 \quad \text{since}\quad [\mathcal{J}_2,\sigma_3]=0\,.\\
\end{split}
\end{equation}
Here $\hat{H}_r$ is the transformed Hamiltonian, $\hat{H}_{m=0}$ is the Hamiltonian $\hat{H}$ with $m$ set equal to zero and $\alpha \in \mathbb{R}$ to guarantee the unitarity of the transformation. Recalling that $\mathcal{J}_2=\frac{\mathcal{J}_+-\mathcal{J}_-}{2i}$, using the commutation relations in \eqref{ec:NCcomutationrelations} and the Hausdorff-Baker-Campbell formula, we get
\begin{equation}
\label{ec:hola}
    \hat{H}_r=\frac{i M}{\sqrt[4]{\kappa\theta}}\left(\mathcal{R}_+-\mathcal{S}_-\right)-\frac{i K}{\sqrt[4]{\kappa\theta}}\left(\mathcal{R}_--\mathcal{S}_+\right)-m\sigma_3\,,
\end{equation}
where we have defined
\begin{equation}\label{ec:defayb}
\begin{array}{c}
    M=\sqrt{1+\sqrt{\kappa\theta}}\left(\sqrt{\kappa}+\omega\sqrt{\theta}\right)\cosh\left(\frac{\alpha}{2}\right)+\sqrt{1-\sqrt{\kappa\theta}}\left(\sqrt{\kappa}-\omega\sqrt{\theta}\right)\sinh\left(\frac{\alpha}{2}\right)  \\
    \,\\
     K=\sqrt{1-\sqrt{\kappa\theta}}\left(\sqrt{\kappa}-\omega\sqrt{\theta}\right)\cosh\left(\frac{\alpha}{2}\right)+\sqrt{1+\sqrt{\kappa\theta}}\left(\sqrt{\kappa}+\omega\sqrt{\theta}\right)\sinh\left(\frac{\alpha}{2}\right) \,.
\end{array}
\end{equation}
Next, we look for the values of $\alpha$ such that either $M=0$ or $K=0$, in order to simplify this expression for $\hat{H}_r$ given in equation \eqref{ec:hola}. Explicitly, we have
\begin{equation}
\label{ec:condayb}
\begin{array}{c}
         M=0 \quad \text{ if and only if }\quad \displaystyle\tanh{\left(\frac{\alpha}{2}\right)}=-\frac{\sqrt{1+\sqrt{\kappa\theta}}}{\sqrt{1-\sqrt{\kappa\theta}}}\cdot\frac{\sqrt{\kappa}+\omega\sqrt{\theta}}{\sqrt{\kappa}-\omega\sqrt{\theta}}  \\
         \,\\
         K=0 \quad  \text{ if and only if }\quad \displaystyle\tanh\left(\frac{\alpha}{2}\right)=-\frac{\sqrt{1-\sqrt{\kappa\theta}}}{\sqrt{1+\sqrt{\kappa\theta}}}\cdot\frac{\sqrt{\kappa}-\omega\sqrt{\theta}}{\sqrt{\kappa}+\omega\sqrt{\theta}} \,.
\end{array}
\end{equation}
In particular, for $\alpha$ to be real, it is necessary and sufficient that

\begin{equation}
\label{ec:alphareal}
    \left|\tanh\left(\frac{\alpha}{2}\right)\right|<1 \,.
\end{equation}
Thus, combining equations \eqref{ec:condayb} and \eqref{ec:alphareal} we get the following conditions

\begin{equation}
\label{ec:hola1}
    \begin{array}{c}
     K=0 \, \text{ and }\,  \alpha \in \mathbb{R} \quad \text{if and only if} \quad \omega^2\theta+2\omega+\kappa>0     \\
     \, \\
     M=0 \, \text{ and }\,  \alpha \in \mathbb{R} \quad \text{if and only if} \quad \omega^2\theta+2\omega+\kappa<0 \,.
    \end{array}
\end{equation}
Note the magnetic field reappears in the expression $\omega^2\theta+2\omega+\kappa=qB$.
%Note that the quantity $2\omega_{\rm NC}:=\omega^2\theta+2\omega+\kappa=qB$ is the physical frequency dependent on the non-commutativity parameters. Moreover, using $\omega = qB/2$ and the expression for $\omega_{\rm NC}$, we can define an effective magnetic field $B_{\rm NC}$ as
%\begin{equation}
%\label{ec:hola2}
%    \omega_{\rm NC} = \frac{q B_{NC}}{2} \quad \text{with} \quad B_{\rm NC}=  \frac{qB^2\theta}{4}+B +\frac{\kappa}{q} \,.
%\end{equation}

As we show below, the Landau problem in the NC phase-space gives the same result as the commutative one as long as one uses the physical magnetic field $B$. If we take $q>0$, the case $K=0$ corresponds to $B>0$, whereas $M=0$ corresponds to $B<0$. Although the Hamiltonian in each of these cases takes a different form, the procedure for solving the spectrum is exactly the same; thus, we only discuss the case $B>0$.  

%Notably, for a fixed value of $q$, $\kappa$ and $\theta$, $\omega_c$ will depend on both the intensity and the direction of the field $\mathbf{B}$ (through $\omega=-qB/2$).

%\begin{figure}[h]
 %   \centering
  %  \includegraphics[scale=0.3]{kappaL.jpg}
   % \caption{$\kappa_L$ as a function of $\omega$ for $\kappa=\theta=0.5$.}
    %\label{fig:kappal}
%\end{figure}

%The idea behind this transformation, may be tracked back to the action of the odd generators on the vectors of the irreducible representations of the graded Lie algebra as described in Appendix \ref{sec:apendice}. In fact, settiang $A=0$ or $B=0$ enables us to diagonalize in a fairly simple manner $\hat{H}_r$.
Taking $\alpha$ such that $K=0$, from equation \eqref{ec:defayb} we may solve for $M$ to get\footnote{In fact, when solving for $M$ we get two possibilities for the expression of the Hamiltonian that differ by a minus sign in the first term, depending on whether $\alpha>0$ or $\alpha<0$. We take $\alpha>0$ given that in both cases the spectrum is the same.}

\begin{equation}
\label{ec:hamiKigualcero}
    \hat{H}_r=i\sqrt{2qB}\left(\mathcal{R}_+-\mathcal{S}_-\right)-m\sigma_3\,.
\end{equation}
We recall that, depending on the value of $\ell=\pm (2k-1)$, the representations of $\mathfrak{sl}(2|1)\oplus \mathfrak{so}(2)$ that intervene are different. Hence we consider, for $\ell=2k-1\geq 0$, in light of the action of the generators $\mathcal{R}_+$ and $\mathcal{S}_-$ (\emph{cf.} Appendix \ref{sec:apendiceaccoddgen}), an eigenvector as a linear combination of
\begin{equation}
    |k,k+n+1,\ell,+\rangle\quad \text{and}\quad \left|k+\tfrac{1}{2},k+\tfrac{1}{2}+n,\ell+1,-\right\rangle \,.
\end{equation}
The eigenvalue equation for the Hamiltonian \eqref{ec:hamiKigualcero}, leads to a system of two linear equations whose solutions are the energy spectrum
\begin{equation}
\label{ec:lmayorlandau}
    E_{\ell\geq 0}^{\pm}
    = \pm\sqrt{2qB(n+1)+m^2}\quad \text{with}\quad n=0,1,2,\dots \,,
\end{equation}
which is independent of $\ell$. Note that the set of vectors $|k,k,\ell,+\rangle$ with $\ell\geq 0$ are eigenvectors of the Hamiltonian associated to the ground-state energy level $E=-m$, which is therefore infinitely degenerate. The fact that there are no states with energy $E=m$ is characteristic to the two dimensional setting, as discussed in \cite{Thaller}. Similarly, for every $n\geq 0$, we have an infinite number of states with the same energy with a different value of $\ell=0,1,2\dots$. On the other hand, if $\ell=-2k+1\leq 0$, we take an  eigenvector as a linear combination of 
\begin{equation}
    |k,k+n,\ell,-\rangle \quad \text{and}\quad  \left|k+\tfrac{1}{2},k+\tfrac{1}{2}+n,\ell-1,+\right\rangle \,.
\end{equation}
Again, the eigenvalue equation for $\hat{H}_r$ leads to a system of two linear equations whose solutions are the energy spectrum for negative angular momentum
\begin{equation}
\label{ec:lmenorlandau}
    E^{\pm}_{\ell\leq 0}=\pm \sqrt{2qB(n-\ell +1)+m^2} \quad \text{with} \quad n=0,1,2,\dots \text{ and }\ell=0,-1,-2 \dots \,.
\end{equation}
We note, in contrast to the positive angular momentum case, that the degeneracy is now finite. In fact, we may put together equations \eqref{ec:lmayorlandau} and \eqref{ec:lmenorlandau} to write the spectrum for arbitrary $\ell$ as

\begin{equation}
\label{ec:finalspectrum}
{E^{+}= \sqrt{2qB\left(n-\tfrac{1}{2}(\ell-|\ell|) +1\right)+m^2}\,,\quad E^-=-\sqrt{2qB\left(n-\tfrac{1}{2}\left(\ell-|\ell|\right)\right)+m^2}}
\end{equation}
with $n=0,1,2\dots$ and $\ell =0,\pm 1, \pm 2,\dots$ . This expression for the spectrum agrees with the well known result in the commutative plane \cite{Thaller} if written in terms of the physical magnetic field $B$. As expected, in the limit $\kappa,\theta\to0$ the spectrum converges smoothly to the commutative one. Furthermore, our result agrees with that of \cite{Yi_2010}, where the Landau problem in the NC phase-space has also been considered. In this case, coordinate NC is achieved via a Moyal-Weyl product of ordinary functions, whereas a Bopp's shift is used to realize momentum NC. 

Finally, it is worth noting that the spectrum described in \eqref{ec:finalspectrum} shows a smooth limit for $\kappa \rightarrow \kappa_c^-$, with $\kappa_c=1/\theta$, as in the non-relativistic case \cite{Falomir1}. Indeed, even though the algebraic structure of the Hamiltonian is very different for $\kappa>\kappa_c$ (whose description following the lines of the article can be performed considering the graded extension of $\frak{su}(2)$), it can be seen that the spectrum of the Landau model presents the same form as in \eqref{ec:finalspectrum} (also with infinite degeneracy), having the same $\kappa \rightarrow \kappa_c^+$ limit. Even more, one can straightforwardly show that the dimensional reduction taking place at the critical point, as mentioned in Section \ref{sec:algebraicstructure}, reduces the Hamiltonian in \eqref{ec:hamiltonianolandau} to that of a harmonic oscillator whose spectrum coincides precisely with both lateral limits for $\kappa \rightarrow \kappa_c$, also with infinite degeneracy this time due to the center of the deformed Heisenberg algebra. Thus, the spectrum of the Landau model in this two-dimensional non-commutative phase space is continuous at $\kappa \theta=1$.

\section{The Cylindrical Well}

\label{sec:cylinder}

In this section we study a Dirac particle in a finite cylindrical well of radius $A$ and depth $V_0$ in the non-commutative plane. In principle, we could add to the Hamiltonian a position dependent step potential $V_0\Theta\left(\hat{\mathbf{X}}^2-A^2\right)$, which corresponds to the timelike component of a vector potential. However, this procedure leads to the Klein paradox, in which a large variation of the potential in a small region of space induces a solution which can only be interpreted in terms of particle creation; as a consequence, one does not obtain the wave function of a single particle confined within the interior of the well \cite{Greiner}. A way out of this problem is to consider a Lorentz scalar potential, instead of a vector potential. This can be done by adding a space-dependent mass term\footnote{Note that a mass term --contrary to a vector potential-- acts in the same way for particles and antiparticles, so no pair creation is to be expected.} 

\begin{equation}
        \hat{H}=\hat{\mathbf{P}}\cdot \boldsymbol{\sigma}-\left(m+V_0\Theta\left(\hat{\mathbf{X}}^2-A^2\right)\right)\sigma_3
\end{equation}
where $V_0$ is a constant with mass units.

%\subsection{The Hamiltonian for $0<\kappa<\kappa_c$} 

In a similar fashion as with the Landau Hamiltonian, we look for a unitary transformation to simplify our problem. From equations \eqref{algeb1} and \eqref{algeb3} we may express $\hat{\mathbf{X}}^2$ in terms of the generators of the graded algebra, and consider the transformation
\begin{equation}
\label{ec:transfunitpozo}
    e^{i\alpha \mathcal{J}_2}\left(\frac{\hat{\mathbf{X}}^2}{\theta}\right)e^{-i\alpha \mathcal{J}_2}=2\mathcal{J}_0-\hat{L}\,.
\end{equation}
with $\tanh(\alpha)=\sqrt{1-\kappa\theta}$, as shown in Appendix C (\emph{cf.} equation \eqref{ec:apendiceirre2}).
%\begin{equation}
%        \frac{\hat{\mathbf{X}}^2}{\theta}=\frac{2}{\sqrt{\kappa\theta}}\mathcal{J}_0+\sqrt{\frac{1}{\kappa \theta}-1}\left(\mathcal{J}_++\mathcal{J}_-\right)-\hat{L}\,.
%\end{equation}
%In any unitary irreducible representation of the non compact group $SL(2,\mathbb{R})$, we may consider the transformation (c.f. Appendix B, \cite{Falomir}), 
%\begin{equation}
 %   e^{i\alpha \mathcal{J}_2}(A\mathcal{J}_0+B\mathcal{J}_1)e^{-i\alpha \mathcal{J}_2}=A\sqrt{1-\frac{B^2}{A^2}}\mathcal{J}_0\,,
%\end{equation}
%where $\mathcal{J}_1=\tfrac{\mathcal{J}_++\mathcal{J_-}}{2}$, and $\tanh (\alpha)=\tfrac{B}{A}$. Thus, letting $A=\tfrac{2}{\sqrt{\kappa \theta}}$ and $B=2\sqrt{\tfrac{1}{\kappa \theta}-1}$, we can diagonalize $\hat{\mathbf{X}}^2$, 
For this choice of $\tanh(\alpha)$ the kinetic term transforms as in the previous section.  
%Recalling the definition $\kappa\theta=\sqrt{\kappa \theta}$, the kinetic term transforms according to,
%\begin{equation}
%\begin{array}{c}
 %   e^{i\alpha \mathcal{J}_2}\left(\hat{\mathbf{P}}\cdot \bm{\sigma}\right)e^{-i\alpha \mathcal{J}_2}= \frac{i\sqrt{\kappa}}{\sqrt{\kappa\theta}}(\mathcal{R}_+-\mathcal{S}_-)\left[\sqrt{1+\kappa\theta}\cosh\left(\frac{\alpha}{2}\right)+\sqrt{1-\kappa\theta}\sinh\left(\frac{\alpha}{2}\right)\right]-
  %  \\
   % \\
%    -\frac{i\sqrt{\kappa}}{\sqrt{\kappa\theta}}(\mathcal{R}_--\mathcal{S}_+)\left[\sqrt{1-\kappa\theta}\cosh\left(\frac{\alpha}{2}\right)+\sqrt{1+\kappa\theta}\sinh\left(\frac{\alpha}{2}\right)\right]   \,.
%\end{array}
%\end{equation}
%where we have used the Hausdorff Baker Campbell formula. We may solve for $\alpha$ using $A$ and $B$ defined above to get\footnote{Actually, we get two possible solutions for $\tanh\left(\frac{\alpha}{2}\right)$. But only this one ensures $\left|\tanh\left(\frac{\alpha}{2}\right)\right|<1$, which is a necessary condition for the transformation to be unitary.} . Putting all together, we get
Taking into account that $[\sigma_3,\mathcal{J}_2]=0$, the transformed Hamiltonian is given by
\begin{equation}
\begin{array}{c}
        \hat{H}_r=e^{i\alpha \mathcal{J}_2}\hat{H}e^{-i\alpha \mathcal{J}_2}=i\sqrt{\frac{2}{\theta}}    \left[(\mathcal{R}_+-\mathcal{S}_-)-\sqrt{1-\kappa\theta}(\mathcal{R}_--\mathcal{S}_+)\right] \\
        \\
 -\left(m+V_0 \Theta\left[\theta\left\{2\mathcal{J}_0-\hat{L}\right\}-A^2\right]\right)\sigma_3\,.
\end{array}
\end{equation}
We now carry out the algebraic solution of the eigenvalue problem. As before, representations will differ depending on the value of $\ell=\pm(2k-1)$, so we start with $\ell=2k-1$. The representation space for the graded Lie algebra for positive $\ell$ is given by $\langle k, \ell,+\rangle \oplus \left\langle k+\tfrac{1}{2},\ell+1,-\right\rangle$ (with the notation in \eqref{ec:horacio1} - \eqref{ec:horacio2}). Hence, we propose a general eigenvector of the form
\begin{equation}
    |\psi\rangle_{\ell\geq 0}= \sum_{n=0}^\infty \alpha_n |k,k+n,\ell,+\rangle + \sum_{n=0}^\infty\beta_n \left|k+\tfrac{1}{2},k+\tfrac{1}{2}+n,\ell+1,-\right\rangle \,.
\end{equation}
Considering the eigenvalue equation $\hat{H}_r|\psi\rangle=E|\psi\rangle$ and taking into account that these vectors are linearly independent, we get a system of coupled three-term recurrence relations for the coefficients $\alpha_n$ and $\beta_n$ given by
\begin{equation}
\label{ec:acopladasrecu}
    \begin{array}{c}
          \left[V_0\Theta\left(\theta\{2n+1\}-A^2\right)+m+E\right]\alpha_n-i\frac{\sqrt{2}}{\sqrt{\theta}} \sqrt{n}\, \beta_{n-1}+\\
          \\
          +i\sqrt{\frac{2}{\theta}} \sqrt{(2k+n)(1-\kappa\theta)}\,\beta_{n}=0\\
         \\
         \left[V_0\Theta\left(\theta\{2n+1\}-A^2\right)+m-E\right]\beta_n-i\sqrt{\frac{2}{\theta}} \sqrt{n+1}\, \alpha_{n+1}+\\
         \\
         +i\sqrt{\frac{2}{\theta}} \sqrt{(2k+n)(1-\kappa\theta)}\,\alpha_{n}=0 \,.\\
    \end{array}
\end{equation}
Due to the presence of the Heaviside function, we define the \textit{matching integer}
\begin{equation}
\label{ec:matchinintegermas}
    n_0^{(+)}=\left[\frac{A^2}{2\theta}-\frac{1}{2}\right] \in \mathbb{N}_0\,,
\end{equation}
with $[\,\cdot\,]:\mathbb{R}\to \mathbb{N}_0$ the floor function, so that \begin{equation}
\begin{cases}
 \begin{array}{c}
      \Theta(\theta(2n+1)-A^2)=0 \quad \text{if} \quad n\leq n_0^{(+)}  \\
      \Theta(\theta(2n+1)-A^2)=1 \quad \text{if} \quad n\geq n_0^{(+)}+1 \,.
 \end{array}
\end{cases}
\end{equation}
We start by simplifying the recurrence relations in equation \eqref{ec:acopladasrecu} with the definitions
\begin{equation}
\label{ec:definicionesss}
    C_n=\frac{\sqrt{(2k+n-1)!}}{\sqrt{n!}}\alpha_n \,, \quad D_n =\frac{\sqrt{(2k+n)!}}{\sqrt{n!}\sqrt{\theta}}\beta_n \, \quad \text{and}\quad b=\sqrt{1-\kappa\theta}\,,
\end{equation}
so that the resulting recurrence relations are given by
\begin{equation}
\label{ec:nmenor}
\begin{array}{c}
(E+m)C_n-i\sqrt{2}(D_{n-1}+bD_n)=0
    \\
    \\
(E-m)\theta D_n-i\sqrt{2}b(2k+n)C_n+i\sqrt{2}(n+1)C_{n+1}=0\\
\end{array}
\end{equation}
if $n\leq n_0^{(+)}$,
while if $n\geq n_0^{(+)}+1$, we get
\begin{equation}
\label{ec:nmayor}
\begin{array}{c}
   (E+m+V_0)C_n-i\sqrt{2}(D_{n-1}+bD_n)=0
    \\
    \\
(E-m-V_0)\theta D_n-i\sqrt{2}b(2k+n)C_n+i\sqrt{2}(n+1)C_{n+1}=0\,.\\
\end{array}
\end{equation}
Equations \eqref{ec:nmenor} may be used to solve for the coefficients $C_n$ with $n\leq n_0^{(+)}+1$ and $D_n$ with $n\leq n_0^{(+)}$, whereas equations \eqref{ec:nmayor} may be used to solve for the coefficients $C_n$ with $n\geq n_0^{(+)}+1$ and $D_n$ with $n\geq n_0^{(+)}$. We note that the coefficients $C_{n_0^{(+)}+1}$ and $D_{n_0^{(+)}}$ may be computed using either pair of equations, a matching condition which will be used later on to determine the spectrum. 

Let us start by considering equation \eqref{ec:nmenor} for $n=0$. This will set the initial condition 
\begin{equation}
\label{ec:condicioninicial}
    (E+m)C_0 - i\sqrt{2}bD_0 = 0\,,
\end{equation}
which allows us to write $D_0$ in terms of $C_0$, and thus, by replacing in the second equation of \eqref{ec:nmenor}, solve for $C_1$. In principle, this procedure could be employed iteratively to solve for the coefficients $C_n$ and $D_n$ up to $n_0^{(+)}+1$ and $n_0^{(+)}$ respectively. Notice that in iteratively solving \eqref{ec:nmayor} for $n\geq n_0^{(+)}+1$ we should not impose an \emph{initial condition} as in \eqref{ec:condicioninicial}, but rather the normalizability of the eigenvector, which demands the coefficient to vanish sufficiently fast.

However, in order to get closed expressions for the coefficients, we will proceed differently. Let us define the generating functions 
\begin{equation}
\label{ec:generatingfunctions}
    F(x):=\sum_{n=0}^{\infty}C_n x^{n}\quad \text{and}\quad G(x):= \sum_{n=0}^{\infty} D_n x^{n} \,.
\end{equation}
We multiply the first equation of \eqref{ec:nmenor} and \eqref{ec:nmayor} by $x^n$ and take the sum from $n=1$. Similarly, we multiply the second equation of \eqref{ec:nmenor} and \eqref{ec:nmayor} by $x^{n+1}$ and sum from $n=0$. Equations \eqref{ec:nmenor} and \eqref{ec:nmayor} thus turn into the set of coupled differential equations
\begin{equation}
\label{ec:diffeqmenor}
\begin{array}{c}
    (E+m)F(x)-i\sqrt{2}(x-b)G(x)=0  \\ \\
    (E-m)\,x\,\theta \, G(x)-i2\sqrt{2}\,k \,b \, x\,F(x)+ix\sqrt{2}(1-bx)F'(x) =0 \,
\end{array}
\end{equation}
and 
\begin{equation}
\label{ec:diffeqmayor}
\begin{array}{c}
    (E+m+V_0)F(x)-i\sqrt{2}(x-b)G(x)=C_0V_0  \\ \\
    (E-m-V_0)\,x\,\theta \, G(x)-i2\sqrt{2}\,k \,b \, x\,F(x)+ix\sqrt{2}(1-bx)F'(x) =0 \,,
\end{array}
\end{equation}
where we have used equation \eqref{ec:condicioninicial} in \eqref{ec:diffeqmenor}. We note that the first equation of \eqref{ec:diffeqmayor} is --besides the different constant coefficient multiplying the first term-- the non-homogeneous version of the first equation of \eqref{ec:diffeqmenor}. The solutions for equation \eqref{ec:diffeqmenor} denoted by $F_L$ and $G_L$, are given by
\begin{equation}
\label{ec:lowww}
\begin{array}{c}
    F_L(x)=(x-b)^{\frac{\theta(m^2-E^2)}{2(b^2-1)}}(1-bx)^{\frac{\theta(E^2-m^2)}{2(b^2-1)}-2k}  \\ \\
    G_L(x)=-{\displaystyle \frac{i(E+m)}{\sqrt{2}}}(x-b)^{-1+\frac{\theta(m^2-E^2)}{2(b^2-1)}}(1-bx)^{\frac{\theta(E^2-m^2)}{2(b^2-1)}-2k}
\end{array}
\end{equation}
while those for equation \eqref{ec:diffeqmayor}, denoted by $F_H$ and $G_H$ are
\begin{equation}
\label{ec:highhh}
\begin{array}{c}
    F_H(x)=\displaystyle \frac{C_0V_0}{E+m+V_0} \,_2F_1 \left(1,2k,1+\frac{\theta(E^2-(m+V_0)^2)}{2(b^2-1)},\frac{b(b-x)}{b^2-1}\right)  \\ \\
    G_H(x)=\displaystyle \frac{iC_0V_0}{\sqrt{2}(b-x)}\left[ \,_2F_1 \left(1,2k,1+\frac{\theta(E^2-(m+V_0)^2)}{2(b^2-1)},\frac{b(b-x)}{b^2-1}\right)-1\right]
\end{array}
\end{equation}
The coefficients, are finally given by
\begin{equation}
\label{ec:finalcoefficients}
 C_n =
\begin{cases}
\begin{array}{c}
     \frac{1}{n!}F_L^{(n)}(0) \quad \text{for }n \leq n_0^{(+)}+1  \\ \\
     \frac{1}{n!}F_H^{(n)}(0) \quad \text{for }n \geq n_0^{(+)}+1
\end{array}
\end{cases}
\quad 
D_n=
\begin{cases}
\begin{array}{c}
     \frac{1}{n!}G_L^{(n)}(0) \quad \text{for }n \leq n_0^{(+)}  \\ \\
     \frac{1}{n!}G_H^{(n)}(0) \quad \text{for }n \geq n_0^{(+)}
\end{array}
\end{cases} \,.
\end{equation}
Now, given the linearity of the recurrence relations, the coefficients obtained by differentiation of the generating functions, are defined up to an $n$-independent scale $R$. So, for the matching conditions, we have
\begin{equation}
\label{ec:r1}
    \frac{R}{(n_0^{(+)}+1)!}F_L^{(n_0^{(+)}+1)}(0)=\frac{1}{(n_0^{(+)}+1)!}F_H^{(n_0^{(+)}+1)}(0) \,.
\end{equation}
and
\begin{equation}
\label{ec:r2}
    \frac{R}{n_0^{(+)}!}G_L^{(n_0^{(+)})}(0)=\frac{1}{n_0^{(+)}!}G_H^{(n_0^{(+)})}(0) \,.
\end{equation}
By eliminating $R$ we get an implicit expression for the energy spectrum,
\begin{equation}
    G_L^{(n_0^{(+)})}(0)F_H^{(n_0^{(+)}+1)}(0)=F_L^{(n_0^{(+)}+1)}(0)G_H^{(n_0^{(+)})}(0)\,,
\end{equation}
with $F_L$ and $G_L$ given in equation \eqref{ec:lowww}, and $F_H$ and $G_H$ in \eqref{ec:highhh}.

We now turn our attention to the negative angular momentum case. The representation space of the graded Lie algebra for negative $\ell$ is given by $\langle k, \ell,-\rangle \oplus \left\langle k+\tfrac{1}{2},\ell-1,+\right\rangle$ so we take a general linear combination of the form
\begin{equation}
    |\psi\rangle_{\ell < 0}= \sum_{n=0}^\infty \alpha_n |k,k+n,\ell,-\rangle + \sum_{n=0}^\infty\beta_n \left|k+\tfrac{1}{2},k+\tfrac{1}{2}+n,\ell-1,+\right\rangle \,.
\end{equation}
From the eigenvalue equation $\hat{H}|\psi\rangle_{\ell \leq 0}=E|\psi\rangle_{\ell \leq 0}$, we get a new set of coupled recurrence relations given by
\begin{equation}
\label{ec:recurr1eleneg}
    \begin{array}{c}
          \left[V_0\Theta\left(\theta\{4k+2n-1\}-A^2\right)+m-E\right]\alpha_n-i\sqrt{\frac{2}{\theta}} \sqrt{2k+n}\, \beta_{n}+ \mbox{}\\
          \\
          \mbox{}+i\sqrt{\frac{2}{\theta}}\sqrt{n} \sqrt{1-\kappa\theta}\,\beta_{n-1}=0\\
         \\
         \left[V_0\Theta\left(\theta\{4k+2n+1\}-A^2\right)+m+E\right]\beta_n-i\sqrt{\frac{2}{\theta}} \sqrt{2k+n}\, \alpha_{n}+ \mbox{}\\
         \\
         \mbox{}+i\sqrt{\frac{2}{\theta}} \sqrt{(n+1)(1-\kappa\theta)}\,\alpha_{n+1}=0 \,.\\
    \end{array}
\end{equation}
Notice that the arguments of the Heaviside step functions is different in this case. We define a new matching integer
\begin{equation}
\label{ec:matchingnlneg}
    n_0^{(-)}=\left[\frac{A^2}{2\theta}-\frac{1}{2}\right]-2k\,,
\end{equation}
so that, together with the definitions given in equation \eqref{ec:definicionesss}, we may write simplified expressions for our recurrence relations \eqref{ec:recurr1eleneg}, 
\begin{equation}
\label{eq:abc1}
\begin{array}{c}
(E-m)C_n-i\sqrt{2}(bD_{n-1}+D_n)=0
    \\
    \\
(E+m)\theta D_n-i\sqrt{2}(2k+n)C_n+i\sqrt{2}b(n+1)C_{n+1}=0\\
\end{array}
\end{equation}
if $n\leq n_0^{(-)}$,
while if $n= n_0^{(-)}+1$ we get
\begin{equation}
\label{ec:ladeldiome}
\begin{array}{c}
(E-m)C_{n_0^{(-)}+1}-i\sqrt{2}(bD_{n_0^{(-)}}+D_{n_0^{(-)}+1})=0
    \\
    \\
(E+m+V_0)\theta D_{n_0^{(-)}+1}-i\sqrt{2}(2k+n_0^{(-)}+1)C_{n_0^{(-)}+1}+i\sqrt{2}b(n_0^{(-)}+2)C_{n_0^{(-)}+2}=0\,,\\
\end{array}
\end{equation}
and finally for $n\geq n_0^{(-)}+2$
\begin{equation}
\label{ec:abc2}
\begin{array}{c}
(E-m-V_0)C_n-i\sqrt{2}(bD_{n-1}+D_n)=0
    \\
    \\
(E+m+V_0)\theta D_n-i\sqrt{2}(2k+n)C_n+i\sqrt{2}b(n+1)C_{n+1}=0 \,.\\
\end{array}
\end{equation}
Next, we proceed similarly to the previous case, by defining generating functions as in \eqref{ec:generatingfunctions}, and rewriting our recurrence relations in terms of $F(x)$ and $G(x)$. Equations \eqref{eq:abc1} and \eqref{ec:abc2} thus become respectively
\begin{equation}
\label{ec:diffeqmenor1}
\begin{array}{c}
    (E-m)F(x)-i\sqrt{2}(bx-1)G(x)=0  \\ \\
    (E+m)\,x\,\theta \, G(x)-i2\sqrt{2}\,k  \, x\,F(x)+ix\sqrt{2}(b-x)F'(x) =0 \,
\end{array}
\end{equation}
and 
\begin{equation}
\label{ec:diffeqmayor1}
\begin{array}{c}
    (E+m+V_0)F(x)-i\sqrt{2}(x-b)G(x)=C_0V_0  \\ \\
    (E-m-V_0)\,x\,\theta \, G(x)-i2\sqrt{2}\,k \,b \, x\,F(x)+ix\sqrt{2}(1-bx)F'(x) =0 \,,
\end{array}
\end{equation}
whose solutions are
\begin{equation}
\label{ec:lowww1}
\begin{array}{c}
    F_L(x)=(x-b)^{-2k+\frac{\theta(m^2-E^2)}{2(b^2-1)}}(1-bx)^{\frac{\theta(E^2-m^2)}{2(b^2-1)}}  \\ \\
    G_L(x)={\displaystyle \frac{i(E-m)}{\sqrt{2}}}(x-b)^{-2k+\frac{\theta(m^2-E^2)}{2(b^2-1)}}(1-bx)^{\frac{\theta(E^2-m^2)}{2(b^2-1)}-1}
\end{array}
\end{equation}
for equation \eqref{ec:diffeqmenor1}, and 
\begin{equation}
\label{ec:highhh1}
\begin{array}{c}
    F_H(x)=\displaystyle \frac{C_0V_0}{E-m-V_0} \,_2F_1 \left(1,2k,1+\frac{\theta(E^2-(m+V_0)^2)}{2(b^2-1)},\frac{bx-1}{b^2-1}\right)  \\ \\
    G_H(x)=-\displaystyle \frac{iC_0V_0}{\sqrt{2}(bx-1)}\left[ \,_2F_1 \left(1,2k,1+\frac{\theta(E^2-(m+V_0)^2)}{2(b^2-1)},\frac{bx-1}{b^2-1}\right)-1\right] \,
\end{array}
\end{equation}
for equation \eqref{ec:diffeqmayor1}. The coefficients may be computed as in \eqref{ec:finalcoefficients}, again up to a global scale $R$. Notice that we have two different ways to compute the coefficient $D_{n_0^{(-)}+1}$: either with the first equation of \eqref{ec:ladeldiome} in terms of $D_{n_0^{(-)}}$ and $C_{n_0^{(-)}+1}$ or with the first equation of \eqref{ec:abc2} in terms of $C_{n_0^{(-)}+2}$ and $D_{n_0^{(-)}+2}$. Similarly, we have two ways of computing $C_{n_0^{(-)}+1}$; either with the second equation of \eqref{eq:abc1} or with the second equation of \eqref{ec:ladeldiome}. Concretely, we have
\begin{equation}
\label{ec:espectrum1}
    Rb\left[(E-m)\frac{F_L^{(n_0^{(-)}+1)}}{{(n_0^{(-)}+1})!}-i\sqrt{2}b\frac{G_L^{(n_0^{(-)})}}{{n_0^{(-)}}!}\right]=(E-m-V_0)\frac{F_H^{(n_0^{(-)}+2)}}{{(n_0^{(-)}+2})!}-i\sqrt{2}\frac{G_H^{(n_0^{(-)}+2)}}{{(n_0^{(-)}+2})!}
\end{equation}

\begin{multline}
\label{ec:espectrum2}
\frac{R(2k+n_0^{(-)}+1)}{b}\left[(E+m)\theta G_L^{(n_0^{(-)})}- i\sqrt{2}(2k+n_0^{(-)})F_L^{(n_0^{(-)})}\right]\\ 
=-\left[(E+m+V_0)\theta G_H^{(n_0^{(-)}+1)}+i\sqrt{2}b F_H^{(n_0^{(-)}+2)}\right]
\end{multline}
Once again, by eliminating $R$, we get an implicit expression for the energy spectrum in terms of derivatives of the generating functions as defined in equations \eqref{ec:lowww1} and \eqref{ec:highhh1}.

\section{Conclusions}

In the framework of quantum mechanics in the non-commutative phase space, we have studied the algebraic structure of fermionic systems described by Dirac Hamiltonians in $(2+1)-$dimensions. 

In so doing, we have considered non-vanishing commutators not only for coordinates but also for momenta, as in Eq. \eqref{ec:conmutconclu}. In this non-commutative phase space we have constructed the generator of rotations, $\hat{L}$, that reduces to its ordinary counterpart in the $\kappa,\theta \to 0$ limit, but is singular at the critical value of the momentum non-commutativity parameter, $\kappa_c = 1/\theta$. At this point, a dimensional reduction takes place, since the system can be described in terms of only one pair of conjugate dynamical variables. This fact requires separate descriptions for the regions $\kappa >\kappa_c$ and $\kappa < \kappa_c$ and thus, these systems present two quantum phases, where only the latter can be connected with the ordinary (commutative) case. For this reason, we have restricted ourselves to the region $\kappa<\kappa_c$. 

In \cite{Falomir1}, where non-relativistic systems were studied from this perspective, it was found that the rotationally invariant Hermitian quadratic forms in the non-commutative dynamical variables generate, in particular, a three-dimensional Lie algebra which corresponds to the non-compact $\mathfrak{sl}(2,\mathbb{R})$. With the goal of extending this method to Dirac-type Hamiltonians, we noticed that the adequate structure to incorporate in the formalism is the graded Lie algebra $\mathfrak{sl}(2|1)$. Indeed, we constructed the set of eight generators of $\mathfrak{sl}(2|1)$ in terms of the non-commutative dynamical variables $\hat{X}_i$, $\hat{P}_i$, $i=1,2$, the Pauli matrices, $\sigma_i$, $i=0,1,2$ and the generator of rotations $\hat{L}$. 
%Notably, the generators are somewhat different depending on whether $\kappa < 0$ or $0 < \kappa < \kappa_c$, and thus, we have considered two non overlapping parameter regions.

Our construction also required the definition of a \textit{total} angular momentum operator $\mathcal{L}$, that together with the generators of the graded Lie algebra were sufficient to express completely the Hamiltonians considered in this article. Thus, the problem associated to computing the spectrum of these Dirac-like Hamiltonians, was translated into the study of the unitary irreducible representations of the graded Lie algebra $\mathfrak{sl}(2|1)\oplus \mathfrak{so}(2)$, where the second factor is generated by $\mathcal{L}$. For completeness, we have included a detailed construction of these \emph{irreps} in a general setting, specifying those which are relevant to our problem. 

%In this way, we have included an explicit construction of these irreducible representations, \textcolor{red}{which we did not find in the literature}. In the same way than in \cite{falomir1}, the positivity of $\hat{X}^2$ (or, equivalently, $\hat{P}^2$), followed from the assumed Hermiticity of the dynamical variables, selected those irreducible representations of $sl(2,\mathbb{R})$ in the discrete class [77, 78] in which the representations are characterized by an integer or half-integer $k$ which bounds from below the eigenvalues of $\mathcal{J}_0$. These were our starting point, to construct the graded representation space, which in the most general case, may be written as the direct sum of the representation space of four different irreducible representations of $sl(2,\mathbb{R})$. For two particular values of $\sigma$, the eigenvalue of $\Sigma$, the generator that together with the generators of $sl(2,\mathbb{R})$ generate the even subalgebra, the number of summands gets reduced to two. As a matter of fact, after computing the quadratic Casimir operator of the graded Lie algebra, we realised that its relation with the dynamical variables, imposes a constraint in which the only values of $\sigma$ with physical significance are precisely these two we have just made reference too. 

From this algebraic point of view, we have analyzed two different problems, namely, the non-commutative extensions of the Landau model and of a cylindrical well. Indeed, taking into account that the characteristic subspaces of these Hamiltonians are unitary irreducible representations of the graded Lie algebra $\mathfrak{sl}(2|1)\oplus \mathfrak{so}(2)$, we obtained in each case an expression for the energy spectrum. 

For the charged particle subject to a uniform magnetic field perpendicular to the NC plane, we found the well-known Landau levels with a constant gap between them proportional to the (covariant) magnetic field.

%We find that it is in good agreement with the one found in the literature \cite{mandal}. We get an infinite degeneracy if both the effective magnetic field and the orbital angular momentum are pointing in the same direction, and a finite degeneracy otherwise. This remark is the key to understanding a very interesting phenomenon that reveals from the interaction between the external magnetic field and the non commutativity parameter between coordinates. For very intense negative magnetic fields, the effective magnetic field eventually \textit{flips} and starts pointing in the same direction than the positive angular momentum. This gives rise to infinitely degenerate levels for particles with $\ell >0$ and a finite degeneracy for particles with $\ell <0$. This inversion of the effective magnetic field comes purely from the coupling between the external magnetic field and the non commutativity paramater between coordinates, and it could result in an experimental method for measuring bounds for non commutativity parameters. 

In the second case, we considered a cylindrical well potential. Since the coordinates are represented by Hermitian operators on a Hilbert space, the boundary of the well is implemented by means of an orthogonal projector onto a subspace of the representation space defined by a condition on the mean square radius. This allows to reduce the eigenvalue equation into a pair of coupled linear three-term recursion for the expansion coefficients of the solution, which includes a potential term that changes when going from the interior to the exterior of the well. This occurs for a particular eigenvalue of $\mathcal{J}_0$, and the equations involving the corresponding coefficient imposes a set of two matching conditions which appear as an effective boundary
condition and provide an implicit expression for the energy spectrum.

\appendix

\section{Adjoint Representation of $\mathfrak{sl}(2|1)$}
\label{sec:apendiceadjunt}
In this appendix we include the matrices corresponding to the generators of $\mathfrak{sl}(2|1)$ in the adjoint representation, the Killing form $\mathbf{K}$ and its inverse $\mathbf{K}^{-1}$. We recall that the graded algebra is spanned by the set of eight generators $\{\mathcal{J}_0,\mathcal{J}_\pm,\Sigma,\mathcal{R}_\pm,\mathcal{S}_\pm\}$. Thus, 

{\tiny

\begin{equation}\label{adj-0}
\begin{array}{c}
      \mathcal{J}^{\rm ad}_-= \left(
\begin{array}{cccccccc}
 0 & 2 & 0 & 0 & 0 & 0 & 0 & 0 \\
 0 & 0 & 0 & 0 & 0 & 0 & 0 & 0 \\
 1 & 0 & 0 & 0 & 0 & 0 & 0 & 0 \\
 0 & 0 & 0 & 0 & 0 & 0 & 0 & 0 \\
 0 & 0 & 0 & 0 & 0 & 0 & 0 & 0 \\
 0 & 0 & 0 & 0 & 1 & 0 & 0 & 0 \\
 0 & 0 & 0 & 0 & 0 & 0 & 0 & 0 \\
 0 & 0 & 0 & 0 & 0 & 0 & 1 & 0 \\
\end{array}
\right)\,,
      
\,\,
\mathcal{J}^{\rm ad}_+=-
  \left(
\begin{array}{cccccccc}
 0 & 0 & 2 & 0 & 0 & 0 & 0 & 0 \\
 1 & 0 & 0 & 0 & 0 & 0 & 0 & 0 \\
 0 & 0 & 0 & 0 & 0 & 0 & 0 & 0 \\
 0 & 0 & 0 & 0 & 0 & 0 & 0 & 0 \\
 0 & 0 & 0 & 0 & 0 & 1 & 0 & 0 \\
 0 & 0 & 0 & 0 & 0 & 0 & 0 & 0 \\
 0 & 0 & 0 & 0 & 0 & 0 & 0 & 1 \\
 0 & 0 & 0 & 0 & 0 & 0 & 0 & 0 \\
\end{array}
\right) \,,
\\ \\
\mathcal{J}^{\rm ad}_0=\left(
\begin{array}{cccccccc}
    0 & 0 & 0 & 0 & 0 & 0 & 0 & 0 \\
  0 & 1 & 0 & 0 & 0 & 0 & 0 & 0 \\
 0 & 0 & -1 & 0 & 0 & 0 & 0 & 0 \\
 0 & 0 & 0 & 0 & 0 & 0 & 0 & 0 \\
  0 & 0 & 0 & 0 & \frac{1}{2} & 0 & 0 & 0 \\
  0 & 0 & 0 & 0 & 0 & -\frac{1}{2} & 0 & 0 \\
  0 & 0 & 0 & 0 & 0 & 0 & \frac{1}{2} & 0 \\
  0 & 0 & 0 & 0 & 0 & 0 & 0 & -\frac{1}{2} \\
\end{array}
\right) \,,
\,\,
\Sigma^{\rm ad}= \left(
\begin{array}{cccccccc}
 0 & 0 & 0 & 0 & 0 & 0 & 0 & 0 \\
 0 & 0 & 0 & 0 & 0 & 0 & 0 & 0 \\
 0 & 0 & 0 & 0 & 0 & 0 & 0 & 0 \\
 0 & 0 & 0 & 0 & 0 & 0 & 0 & 0 \\
 0 & 0 & 0 & 0 & 1 & 0 & 0 & 0 \\
 0 & 0 & 0 & 0 & 0 & 1 & 0 & 0 \\
 0 & 0 & 0 & 0 & 0 & 0 & -1 & 0 \\
 0 & 0 & 0 & 0 & 0 & 0 & 0 & -1 \\
\end{array}
\right) \,,

\\ \\
\mathcal{R}^{\rm ad}_+=\left(
\begin{array}{cccccccc}
 0 & 0 & 0 & 0 & 0 & 0 & 0 & 1 \\
 0 & 0 & 0 & 0 & 0 & 0 & 1 & 0 \\
 0 & 0 & 0 & 0 & 0 & 0 & 0 & 0 \\
 0 & 0 & 0 & 0 & 0 & 0 & 0 & -\frac{1}{2} \\
 -\frac{1}{2} & 0 & 0 & -1 & 0 & 0 & 0 & 0 \\
 0 & 0 & -1 & 0 & 0 &
   0 & 0 & 0 \\
 0 & 0 & 0 & 0 & 0 & 0 & 0 & 0 \\
 0 & 0 & 0 & 0 & 0 & 0 & 0 & 0 \\
\end{array}
\right) \,,
     \, \,
\mathcal{R}^{\rm ad}_-= \left(
\begin{array}{cccccccc}
 0 & 0 & 0 & 0 & 0 & 0 & 1 & 0 \\
 0 & 0 & 0 & 0 & 0 & 0 & 0 & 0 \\
 0 & 0 & 0 & 0 & 0 & 0 & 0 & 1 \\
 0 & 0 & 0 & 0 & 0 & 0 & \frac{1}{2} & 0 \\
 0 & 1 & 0 & 0 & 0 & 0
   & 0 & 0 \\
 \frac{1}{2} & 0 & 0 & -1 & 0 & 0 & 0 & 0 \\
 0 & 0 & 0 & 0 & 0 & 0 & 0 & 0 \\
 0 & 0 & 0 & 0 & 0 & 0 & 0 & 0 \\
\end{array}
\right) \,,
    \\ \\
\mathcal{S}^{\rm ad}_+=\left(
\begin{array}{cccccccc}
 0 & 0 & 0 & 0 & 0 & 1 & 0 & 0 \\
 0 & 0 & 0 & 0 & 1 & 0 & 0 & 0 \\
 0 & 0 & 0 & 0 & 0 & 0 & 0 & 0 \\
 0 & 0 & 0 & 0 & 0 & \frac{1}{2} & 0 & 0 \\
 0 & 0 & 0 & 0 & 0 & 0 & 0 & 0 \\
 0 & 0 & 0 & 0 & 0 & 0 & 0 & 0 \\
 -\frac{1}{2} & 0 & 0 & 1 & 0 & 0 & 0 & 0 \\
 0 & 0 & -1 & 0 & 0 &
   0 & 0 & 0 \\
\end{array}
\right) \,,
     \, \,
\mathcal{S}^{\rm ad}_-=\left(
\begin{array}{cccccccc}
 0 & 0 & 0 & 0 & 1 & 0 & 0 & 0 \\
 0 & 0 & 0 & 0 & 0 & 0 & 0 & 0 \\
 0 & 0 & 0 & 0 & 0 & 1 & 0 & 0 \\
 0 & 0 & 0 & 0 & -\frac{1}{2} & 0 & 0 & 0 \\
 0 & 0 & 0 & 0 & 0 & 0 & 0 & 0 \\
 0 & 0 & 0 & 0 & 0 & 0 & 0 & 0 \\
 0 & 1 & 0 & 0 & 0 & 0
   & 0 & 0 \\
 \frac{1}{2} & 0 & 0 & 1 & 0 & 0 & 0 & 0 \\
\end{array}
\right) \,,
\\ \\

\mathbf{K}=\left(
\begin{array}{cccccccc}
 1 & 0 & 0 & 0 & 0 & 0 & 0 & 0 \\
 0 & 0 & -2 & 0 & 0 & 0 & 0 & 0 \\
 0 & -2 & 0 & 0 & 0 & 0 & 0 & 0 \\
 0 & 0 & 0 & -4 & 0 & 0 & 0 & 0 \\
 0 & 0 & 0 & 0 & 0 & 0 & 0 & 2 \\
 0 & 0 & 0 & 0 & 0 & 0 & -2 & 0 \\
 0 & 0 & 0 & 0 & 0 & 2 & 0 & 0 \\
 0 & 0 & 0 & 0 & -2 & 0 & 0 & 0 \\
\end{array}
\right) \,,
\,\,
\mathbf{K}^{-1}=-\left(
\begin{array}{cccccccc}
 -1 & 0 & 0 & 0 & 0 & 0 & 0 & 0 \\
 0 & 0 & \frac{1}{2} & 0 & 0 & 0 & 0 & 0 \\
 0 & \frac{1}{2} & 0 & 0 & 0 & 0 & 0 & 0 \\
 0 & 0 & 0 & \frac{1}{4} & 0 & 0 & 0 & 0 \\
 0 & 0 & 0 & 0 & 0 & 0 & 0 & \frac{1}{2} \\
 0 & 0 & 0 & 0 & 0 & 0 & -\frac{1}{2} & 0 \\
 0 & 0 & 0 & 0 & 0 & \frac{1}{2} & 0 & 0 \\
 0 & 0 & 0 & 0 & -\frac{1}{2} & 0 & 0 & 0 \\
\end{array}
\right) \,,

   \end{array}
\end{equation}

}

\section{Action of the Odd Generators}
\label{sec:apendiceaccoddgen}

Here we summarize the action of the odd generators on vectors of the representation space for $\sigma = \pm 2k.$ 

\subsection{Case $\sigma=2k$} 

\begin{equation}
    \begin{array}{c}
         \mathcal{R}_+|k,k+n,\sigma\rangle =  0\\
         \\
         \mathcal{R}_-|k,k+n,\sigma\rangle =  0
    \end{array}
\end{equation}

\begin{equation}
    \begin{array}{c}
         \mathcal{S}_+|k,k+n,\sigma\rangle =\sqrt{2k+n}\left|k+\tfrac{1}{2},k+\tfrac{1}{2}+n,\sigma-1\right\rangle\\       
         \\
         \mathcal{S}_-|k,k+n,\sigma\rangle =\sqrt{n}\left|k+\tfrac{1}{2},k+\tfrac{1}{2}+(n-1),\sigma-1\right\rangle\\ 
         \\
    \end{array}
\end{equation}

\begin{equation}
    \begin{array}{c}
         \mathcal{R}_+\left|k+\tfrac{1}{2},k+\tfrac{1}{2}+n,\sigma-1\right\rangle =  \sqrt{n+1}|k,k+n+1,\sigma\rangle\\
         \\
         \mathcal{R}_-\left|k+\tfrac{1}{2},k+\tfrac{1}{2}+n,\sigma-1\right\rangle = \sqrt{2k+n}|k,k+n,\sigma \rangle
    \end{array}
\end{equation}
\begin{equation}
    \begin{array}{c}
         \mathcal{S}_+\left|k+\tfrac{1}{2},k+\tfrac{1}{2}+n,\sigma-1\right\rangle =0\\       
         \\
         \mathcal{S}_-\left|k+\tfrac{1}{2},k+\tfrac{1}{2}+n,\sigma-1\right\rangle =0\\ \\
    \end{array}
\end{equation}

\subsection{Case $\sigma=-2k$}

\begin{equation}
    \begin{array}{c}
         \mathcal{R}_+|k,k+n,\sigma\rangle =  \sqrt{2k+n}\left|k+\tfrac{1}{2},k+n+\tfrac{1}{2},\sigma+1\right\rangle\\
         \\
         \mathcal{R}_-|k,k+n,\sigma\rangle =  \sqrt{n}\left|k+\tfrac{1}{2},k+\tfrac{1}{2}+(n-1),\sigma+1\right\rangle
    \end{array}
\end{equation}
\begin{equation}
    \begin{array}{c}
         \mathcal{S}_+|k,k+n,\sigma\rangle =0\\       
         \\
         \mathcal{S}_-|k,k+n,\sigma\rangle =0\\ 
         \\
    \end{array}
\end{equation}

\begin{equation}
    \begin{array}{c}
         \mathcal{R}_+\left|k+\tfrac{1}{2},k+\tfrac{1}{2}+n,\sigma+1\right\rangle =  0\\
         \\
         \mathcal{R}_-\left|k+\tfrac{1}{2},k+\tfrac{1}{2}+n,\sigma+1\right\rangle =  0
    \end{array}
\end{equation}
\begin{equation}
    \begin{array}{c}
         \mathcal{S}_+\left|k+\tfrac{1}{2},k+\tfrac{1}{2}+n,\sigma+1\right\rangle =\sqrt{n+1}\left|k,k+n+1,\sigma\right\rangle\\
         \\
         \mathcal{S}_-\left|k+\tfrac{1}{2},k+\tfrac{1}{2}+n,\sigma+1\right\rangle =\sqrt{2k+n}\left|k,k+n,\sigma\right\rangle\\ 
         \\
    \end{array}
\end{equation}

\section{Unitary Transformations in $\mathfrak{sl}(2,\mathbb{R})$}

The generators in the fundamental (nonunitary irreducible)
representation of $\mathfrak{sl}(2,\mathbb{R})$ can be chosen as \cite{falomir2}

\begin{equation}
    X_0=-\frac{1}{2}\sigma_2\,\quad  X_1=\frac{i}{2}\sigma_1\,\quad  X_2=\frac{i}{2}\sigma_3\,,
\end{equation}
where $\sigma_i$, $i=1,2,3$ are the Pauli matrices. These generators satisfy the commutation relations

\begin{equation}
    [X_\mu,X_\nu]=-i\epsilon_{\mu\nu\lambda}X^{\lambda} \,
\end{equation}
where $X^{\mu}=\eta^{\mu\nu}X_\nu$ with $\eta^{\mu\nu}={\rm diag}(1,-1,-1)$. It is straightforward to verify by direct computation that

\begin{equation}
    e^{i \alpha X_{2}}\left(A X_{0}+B X_{1}\right) e^{-i \alpha X_{2}}=A \sqrt{1-\frac{B^{2}}{A^{2}}} X_{0}
\end{equation}
for $A,B\in \mathbb{R}$ with $|A|>|B|$, if we take $\tanh(\alpha)=B/A$. Then, in any unitary representation of $\mathfrak{sl}(2,\mathbb{R})$ (with Hermitian generators $\mathcal{J}_i$)
we also have

\begin{equation}
\label{ec:apendiceirre2}
    e^{i \alpha \mathcal{J}_{2}}\left(A \mathcal{J}_{0}+B \mathcal{J}_{1}\right) e^{-i \alpha \mathcal{J}_{2}}=A \sqrt{1-\frac{B^{2}}{A^{2}}} \mathcal{J}_{0}\,.
\end{equation}
Indeed, the left hand side of Eq. \eqref{ec:apendiceirre2} can be formally written as

\begin{equation}
    \begin{array}{c}
         \displaystyle\sum_{k=0}^{\infty} \frac{(i \alpha)^{k}}{k !}\left(\left[\mathcal{J}_{2}, \cdot \,\right]^{k}\left(A \mathcal{J}_{0}+B \mathcal{J}_{1}\right)\right) \\
         \\
         =(A \cosh \alpha-B \sinh \alpha) \mathcal{J}_{0}+(B \cosh \alpha-A \sinh \alpha) \mathcal{J}_{1}
    \end{array}
\end{equation}
The coefficient of $\mathcal{J}_1$ on the right hand side of this equation vanishes if we choose $\tanh(\alpha) = B/A$ (for $|B/A| < 1$), in which case one gets the right hand side of Eq.
\eqref{ec:apendiceirre2}.

Similarly, 

\begin{equation}
    \begin{array}{c}
e^{i \beta \mathcal{J}_{1}}\left(A \mathcal{J}_{0}+B \mathcal{J}_{2}\right) e^{-i \beta \mathcal{J}_{1}}= \\
\\
=(A \cosh \beta+B \sinh \beta) \mathcal{J}_{0}+(A \sinh \beta+B \cosh \beta) \mathcal{J}_{2}\,.
\end{array}
\end{equation}
Therefore, by choosing $\tanh(\beta)=-B/A$, for real $A,B$ with $|A|>|B|$, one gets
\begin{equation}
\label{ec:apendiceirre1}
    e^{i \beta \mathcal{J}_{1}}\left(A \mathcal{J}_{0}+B \mathcal{J}_{2}\right) e^{-i \beta \mathcal{J}_{1}}=A \sqrt{1-\frac{B^{2}}{A^{2}}} \mathcal{J}_{0}\,.
\end{equation}

\bigskip

\noindent{\bf Acknowledgments:}

This work was supported by CONICET, and UNLP under project X909. 

\sloppy

\printbibliography[title=\textsc{References}]

% --------------------------------------------------------------
\end{document}